\documentclass[twocolumn,nofootinbib,superscriptaddress,preprintnumbers,floatfix]{revtex4}

\usepackage[bookmarks=false,hyperfootnotes=false]{hyperref}
\usepackage{graphicx}
\usepackage{amsmath}
\usepackage{xspace}
\usepackage{color}

\newcommand{\eq}[1]{Eq.~\eqref{eq:#1}}
\newcommand{\eqs}[2]{Eqs.~\eqref{eq:#1} and \eqref{eq:#2}}
\renewcommand{\sec}[1]{Sec.~\ref{sec:#1}}
\newcommand{\secs}[2]{Secs.~\ref{sec:#1} and \ref{sec:#2}}
\newcommand{\subsec}[1]{Sec.~\ref{subsec:#1}}
\newcommand{\fig}[1]{Fig.~\ref{fig:#1}}
\newcommand{\app}[1]{Appendix~\ref{app:#1}}

\newcommand{\abs}[1]{\lvert#1\rvert}
\newcommand{\ord}[1]{\mathcal{O}(#1)}

\newcommand{\Mae}[3]{\bigl\langle#1\bigl\lvert#2\bigr\rvert#3\bigr\rangle}

\newcommand{\be}{\begin{equation}}
\newcommand{\ee}{\end{equation}}

\newcommand{\df}{\mathrm{d}}

\newcommand{\Tau}{\mathcal{T}}
\newcommand{\e}{\epsilon}

\newcommand{\bn}{\bar{n}}

\newcommand{\cA}{\mathcal{A}}
\newcommand{\cL}{\mathcal{L}}
\newcommand{\cM}{\mathcal{M}}
\newcommand{\cO}{\mathcal{O}}
\newcommand{\cI}{\mathcal{I}}

\newcommand{\GeV}{\,\mathrm{GeV}}
\newcommand{\TeV}{\,\mathrm{TeV}}

\newcommand{\nn}{\nonumber}

\newcommand{\as}{\alpha_s}
\newcommand{\Tauj}{\Tau_j}
\newcommand{\Taucut}{\Tau^\mathrm{cut}}
\newcommand{\pTj}{p_{Tj}}
\newcommand{\pTcut}{p_T^\mathrm{cut}}
\newcommand{\Ecm}{E_\mathrm{cm}}
\newcommand{\Gcusp}{\Gamma_{\rm cusp}^g}

\newcommand{\cut}{\mathrm{cut}}
\newcommand{\jet}{\mathrm{jet}}

\newcommand{\kt}{k_\mathrm{T}}
\newcommand{\SCETI}{SCET$_{\rm I}$\xspace}
\newcommand{\SCETII}{SCET$_{\rm II}$\xspace}

\allowdisplaybreaks[4]

\begin{document}


\preprint{\vbox{\hbox{DESY 12-104}}}

\title{Resummation Properties of Jet Vetoes at the LHC}

\author{Frank J.~Tackmann}
\affiliation{Theory Group, Deutsches Elektronen-Synchrotron (DESY), D-22607 Hamburg, Germany\vspace{0.5ex}}

\author{Jonathan R.~Walsh}
\affiliation{Ernest Orlando Lawrence Berkeley National Laboratory,
University of California, Berkeley, CA 94720\vspace{0.5ex}}

\author{Saba Zuberi}
\affiliation{Ernest Orlando Lawrence Berkeley National Laboratory,
University of California, Berkeley, CA 94720\vspace{0.5ex}}

\date{June 19, 2012}

\begin{abstract}

Jet vetoes play an important role at the LHC in the search for the Higgs and ultimately in precise measurements of its properties. Many Higgs analyses divide the cross section into exclusive jet bins to maximize the sensitivity in different production and decay channels. For a given jet category, the veto on additional jets introduces sensitivity to soft and collinear emissions, which causes logarithms in the perturbative expansion that need to be resummed to obtain precise predictions. We study the higher-order resummation properties of several conceptually distinct kinematic variables that can be used to veto jets in hadronic collisions. We consider two inclusive variables, the scalar sum over $p_T$ and beam thrust, and two corresponding exclusive variables based on jet algorithms, namely the largest $p_T$ and largest beam thrust of a jet. The inclusive variables can, in principle, be resummed to higher orders. We show that for the jet-based variables, there are dual effects due to clustering in the jet algorithm for both large and small jet radius $R$ that make a complete resummation at or beyond next-to-leading logarithmic order (NLL) challenging. For $R\sim 1$, the clustering of soft and collinear emissions gives $\ord{1}$ contributions starting at next-to-next-to-leading logarithmic order that are not reproduced by an all-orders soft-collinear factorization formula and therefore are not automatically resummed by it. For $R \ll 1$, clustering induces logarithms of $R$ that contribute at NLL in the exponent of the cross section, which cannot be resummed with currently available methods. We explicitly compute the leading jet clustering effects at $\ord{\alpha_s^2}$ and comment on their numerical size.

\end{abstract}

\maketitle

\section{Introduction}
\label{sec:intro}

The LHC is at the brink of discovering (or ruling out) a Standard-Model-like Higgs boson~\cite{ATLAS:2012ae, Chatrchyan:2012tx}. Many Higgs and new-physics analyses divide the data into categories based on the number of hard jets in the final state. This ``jet-binning'' significantly enhances the experimental sensitivity to various combinations of Higgs production and decay channels. The experimental distinction between different channels is of course an important requirement to determine the Higgs properties. Any coupling measurement requires precise and reliable theory predictions of the jet cross sections used in the experimental analysis.

By vetoing jets, one restricts the phase space for additional emissions. This makes the cross section sensitive to soft and collinear radiation, which induces Sudakov double logarithms of the jet-veto variable at each order in perturbation theory. At small enough cuts the logarithmic corrections become large and dominate the perturbative series, which degrades the reliability of fixed-order perturbation theory. To obtain precise predictions and robust uncertainty estimates it is important to understand the structure of large logarithms and ideally resum them to all orders in perturbation theory. The form of the jet-veto logarithms and the precise structure of the logarithmic series depends on the details of how the jet veto is imposed.

In this paper we consider four kinematic variables that are conceptually and theoretically distinct, and study the prospects for resumming the respective large logarithms when these (types of) variables are used to veto central jets. For definiteness we will discuss the process $gg\to H + 0$ jets in the following, but our analysis applies to the production of any color-singlet in conjunction with a veto on central jets. It also carries over to generic processes where one requires $N$ signal jets in the final state and places a veto on additional jets, i.e. when measuring an exclusive $N$-jet cross section.

In principle, one can think of various kinematic variables that can be employed to enforce a veto on hard emissions. The four variables we will discuss are summarized in Table~\ref{table:variables} and are described next. They are classified as either inclusive or exclusive (jet-based) variables and according to their sensitivity to either the virtuality or $p_T$ of emissions.

The inclusive variables we consider are beam thrust~\cite{Stewart:2009yx} and the scalar sum of $p_T$,
\begin{align} \label{eq:inclvarTB}
\Tau_B &= \sum_m \abs{\vec{p}_{Tm}}\,e^{-\abs{y_m - Y}}
\,,\\
E_T &= \sum_m \abs{\vec{p}_{Tm}}\label{eq:inclvarET}
\,,\end{align}
where $\vec{p}_{Tm}$ and $y_m$ are the transverse momenta and rapidities with respect to the beam axis of all particles in the final state, but excluding the Higgs decay products, and $Y$ is the rapidity of the Higgs boson.%
\footnote{For cases like $H\to WW$ where $Y$ cannot be measured directly, one can also consider the analog of $\Tau_B$ defined in the hadronic center-of-mass frame by setting $Y = 0$ in \eq{inclvarTB}.}
For these inclusive variables one sums over all hadrons $m$ in the final state, which at the perturbative level corresponds to constraining the sum over all emissions. As a result such inclusive variables are theoretically the cleanest and best understood. The complete resummation of logarithms at small $\Tau_B$ and $E_T$ has been carried out to next-to-next-to-leading logarithmic order (NNLL)~\cite{Stewart:2010pd, Berger:2010xi} and next-to-leading logarithmic order (NLL)~\cite{Papaefstathiou:2010bw}, respectively, and the extension to higher logarithmic orders poses no conceptual difficulties.  The inclusive nature of these variables makes it challenging to measure them in a hadron-collider environment. These experimental issues can be mitigated by summing over jets rather than hadrons, as is done in current event-shape measurements at the LHC~\cite{Khachatryan:2011dx, :2012np}, or by only summing over charged tracks, which can then be corrected for in the measurement.

The difference between $\Tau_B$ and $E_T$ is in the different rapidity weighting of emissions. In the rest frame of the Higgs, where $Y = 0$, $\Tau_B$ is equivalent to the sum over the small light-cone component of momenta, $\Tau_B = \sum_k (E_k - \abs{p_k^z}) \sim t/m_H$, where $t$ is the spacelike virtuality of the colliding hard partons after initial-state radiation. Hence, by measuring beam thrust $\Tau_B$ one is sensitive to the virtuality scale of emissions. In contrast, by measuring $E_T$ one is sensitive to the $p_T$ scale of emissions, and one might call it ``beam broadening''. This difference in sensitivity to virtuality vs.~$p_T$ causes the logarithmic series at small $\Tau_B$ or $E_T$ to have very different structures.  In the parton shower, this difference is analogous to the different Sudakov form factors for virtuality and $p_T$ ordered showers.

The exclusive jet-based variables we consider are the largest beam-thrust or $p_T$ of a jet, given by
\begin{align} \label{eq:exclvarTau}
\Tauj &= \max_{m\in j(R)}\, \abs{\vec{p}_{Tm}}\,e^{-\abs{y_m - Y}}
\,,\\
\label{eq:exclvarpT}
\pTj &= \max_{m\in j(R)}\, \abs{\vec{p}_{Tm}}
\,,\end{align}
where $\vec{p}_{Tj}$ and $y_j$ are the jet's transverse momentum and rapidity. They have the same sensitivity to the virtuality or $p_T$ scale of emissions as their inclusive counterparts $\Tau_B$ and $E_T$ in \eqs{inclvarTB}{inclvarET}. The exclusive variables are based on identifying jets $j(R)$ of size $R$ and considering the largest contribution from a jet. Perturbatively, this corresponds to constraining the maximum of all emissions at a typical ``resolution scale'' $R$, which is in contrast to constraining the sum of emissions as in the inclusive variables. We focus on jets using the $\kt$ class of algorithms \cite{Catani:1991hj, Catani:1993hr, Ellis:1993tq, Dokshitzer:1997in, Cacciari:2008gp}. The jet-based variables are more straightforward to use experimentally, and a $p_T$ veto on jets using the anti-$\kt$ algorithm is the common choice in experiments. On the other hand, the variable's inherent dependence on the jet algorithm and the resolution scale $R$ make them considerably less tractable theoretically. The resummation for $\pTj$ was considered recently in Refs.~\cite{Banfi:2012yh, Becher:2012qa}.

\begin{table}
\begin{tabular}{c|cc}
\hline\hline
& \multicolumn{2}{c}{sensitive to}
\\
& virtuality & $p_T$ \\ \hline
inclusive & $\Tau_B$ & $E_T$ \\
jet-based (exclusive) & $\Tauj$ & $\pTj$ \\
\hline
constraint & $\leq\Taucut$ & $\leq\pTcut$ \\
\hline\hline
\end{tabular}
\caption{Classification of different jet-veto variables.
\label{table:variables}}
\end{table}

We implement the veto on central jets by putting a constraint
\begin{align}
\Tau_B &\leq \Taucut
\,,&
E_T &\leq \pTcut
\,,\nn\\
\Tauj &\leq \Taucut
\,,&
\pTj &\leq \pTcut
\,,\end{align}
and define the small parameter $\lambda$ by
\begin{equation}
\lambda^2 = \Taucut/m_H
\qquad\text{or}\qquad
\lambda = \pTcut/m_H
\,.\end{equation}
The cross section with a $\Taucut$ or $\pTcut$ veto contains Sudakov logarithms $\alpha_s^n \ln^m \lambda$ with $m \leq 2n$. In the region of small $\lambda$ the logarithms dominate the cross section and we want to resum them. At the same time, any nonlogarithmic contributions that depend on $\Taucut$ or $\pTcut$ are suppressed by relative powers of $\lambda$, and can be added to the resummed result.

For the jet-based variables, one can distinguish two cases of how the jet size $R$ is counted relative to $\lambda$, we can consider either $R \sim \lambda$ or $R \gg \lambda$.%
\footnote{This counting is a natural proxy for the two general cases of $R$ either scaling as a positive power of $\lambda$ or $R$ not scaling with $\lambda$.  The difference amounts to whether we formally count $\ord{R}$ terms as power corrections in $\lambda$ or not and logarithms of $R$ as $\ln\lambda$ or not.}
For this distinction it is irrelevant whether one considers $R \sim 1$ or $R \ll 1$. We stress that which of these two formal cases is in the end more appropriate in practice, for given numerical values of $R$ and $\Taucut/m_H$ or $\pTcut/m_H$, is a separate question that needs to be studied numerically and will be addressed in \sec{conclusions}.

As we will show in this paper, there are competing effects in either limit arising from the dependence on the jet algorithm, which spoil the complete logarithmic resummation.  Jet algorithm dependent effects first arise at $\ord{\as^2}$ in the form
\begin{equation} \label{eq:twoloopterm}
\as^2\, f^{(2)}_{\rm alg} (R)\, \ln \lambda
\,.\end{equation}
The function $f^{(2)}_{\rm alg}(R)$ depends on the algorithm and the jet-veto variable, and in general contains terms of $\ord{R^2}$, constant terms, and $\ln R$ terms. These were calculated in Ref.~\cite{Banfi:2012yh} for the $\pTj$ veto.

For $R\gg\lambda$ the jet algorithm mixes soft and collinear contributions in the measurement, giving rise to a term in \eq{twoloopterm} of the form
\begin{equation}
\alpha_s^2 R^2 \ln\lambda \,,
\end{equation}
at leading order in $R$. When keeping the full $R$ dependence, these soft-collinear mixing terms inhibit an all-order soft-collinear factorization in the measurement at leading power in $\lambda$. In Ref.~\cite{Becher:2012qa}, a factorization formula for the $\pTj$ veto was presented in the limit $\lambda \ll R \sim 1$. Their derivation does not account for the effect of these mixing terms to all orders, effectively assuming that they are power-suppressed, and hence breaks down for $R\sim1$.

For $R \sim \lambda$, the mixing terms inhibiting the soft-collinear factorization of the measurement can be regarded as power corrections in $\lambda$, so factorization goes through. However, in this limit clustering effects that change the boundary of the jet at each order in $\as$ introduce corrections that depend on $\ln R$. These are referred to as clustering logarithms and were first pointed out in Ref.~\cite{Banfi:2005gj} and were studied in Refs.~\cite{Delenda:2006nf, KhelifaKerfa:2011zu, Hornig:2011tg, Kelley:2012kj, Kelley:2012zs}. The clustering logarithms in $f^{(2)}_{\rm alg}(R)$ give a contribution of the form
\begin{equation}
\alpha_s^2\, \ln R\, \ln\lambda \,.
\end{equation}
They are distinct from those previously studied since they are associated with collinear rather than soft divergences within each jet. The $\ord{\as^2}$ term in \eq{twoloopterm} is the first in an all-orders series of terms of the form
\begin{equation}
\alpha_s^n\, f^{(n)}_{\rm alg} (R)\, \ln\lambda \,, \qquad f^{(n)}_{\rm alg} (R) \supset \ln^k R \,, \;\;  k \le n-1 \,.
\end{equation}
For $R\sim\lambda$ we have to count $\ln R\sim\ln\lambda$, so these clustering logarithms give a new contribution at NLL at each higher order in $\alpha_s$. Hence, they spoil the complete logarithmic resummation at NLL and beyond.

At a formal level, this means one is stuck between a rock and a hard place. We would like to consider $R$ as small to justify not resumming soft-collinear mixing terms, by treating them as unresummed power corrections. But at the same time we would like to treat $R$ as large, to avoid having to count $\ln R$ terms as large logarithms.

A possible way forward would be to resum those logarithms whose all-order series is known while simultaneously quantifying the effect of the terms that cannot be resummed.  If these effects can be appropriately folded into uncertainty estimates, then a reliable theoretical prediction can be obtained.  The formal power counting is less important in this case, and effectively the two formal limits for $R$ are unified by reflecting the important terms from each case in the perturbative uncertainties.

The remainder of this paper is organized as follows. In \sec{factorization} we present an overview of the necessary steps to achieve soft-collinear factorization and resummation, focusing on the properties of the measurement.  We present the factorization formulas for the inclusive observables in \subsec{inclobservables} and for the exclusive observables in \subsec{exclobservables}. In \sec{scmixing} we discuss the soft-collinear mixing in detail and show that it gives an $\ord{R^2}$ contribution to the rate. We demonstrate this by power counting as well as explicit calculation of the mixing terms at $\ord{\as^2}$, the details of which are given in \app{scmixingcalc}. In \sec{clusteringlogs} we discuss the clustering logarithms in detail and calculate their contribution at $\ord{\as^2}$, with the details of the calculation given in \app{clusteringlogcalc}. We use non-Abelian exponentiation and the collinear sensitivity of the clustering logarithms to show that their contribution to the cross section exponentiates and for $R\sim \lambda$ contributes at NLL in the exponent. In \sec{conclusions} we summarize our findings and give an outlook based on a numerical analysis of the size of the $\ord{\as^2}$ mixing and clustering terms for different values of $R$.

\section{Factorization Formulas}
\label{sec:factorization}

The large Sudakov logarithms in the jet-vetoed cross sections arise as a remnant of the cancellation of soft and collinear IR divergences between virtual and real contributions. Their resummation is thus intimately tied to the universal structure of QCD amplitudes in the soft and collinear limit. Furthermore, resummation relies on the fact that soft gluon emissions from energetic particles are eikonal and that the total soft eikonal matrix element factorizes from the remaining amplitude.

A convenient framework to study the logarithmic structure of the cross section is provided by soft-collinear effective theory (SCET)~\cite{Bauer:2000ew, Bauer:2000yr, Bauer:2001ct, Bauer:2001yt, Bauer:2002nz, Beneke:2002ph}, which makes the soft-collinear limit of QCD manifest at a Lagrangian and operator level using a systematic power expansion in the small parameter $\lambda$. The resummation of Sudakov logarithms is then achieved by standard effective-theory methods through a systematic scale separation and renormalization group evolution between the scales.
For a detailed discussion of this procedure in the context of SCET we refer the reader to the literature. In the following, we give a schematic overview of the basic steps, concentrating on the features that are important for our further analysis.

After matching full QCD onto SCET, the cross section in SCET for $gg\to H$ with no hard jets in the final states has the schematic form (for details see e.g. Refs.~\cite{Stewart:2009yx, Berger:2010xi})
\begin{equation} \label{eq:dsigma}
\df\sigma_{gg\to H} \sim \lvert C_{ggH}(\mu) \rvert^2 \Mae{p_ap_b}{\cO_{ggH}(\mu)^{\dagger} \widehat{\cM} \, \cO_{ggH}(\mu) }{p_ap_b}
\,.\end{equation}
The incoming (anti)protons have momenta
\begin{equation}
p_{a,b}^\mu = \Ecm \frac{n_{a,b}^\mu}{2}
\quad\text{with}\quad
n_a^\mu = (1, \hat z)
\,,\quad
n_b^\mu = (1, -\hat z)
\,.\end{equation}
The Wilson coefficient $C_{ggH}(\mu)$ arises from matching onto the $\cO_{ggH}(\mu)$ operator in SCET and encodes the hard-scattering contributions, which live at the hard scale $\mu_H \sim m_H$. The measurement operator $\widehat{\cM}$ implements the phase-space cuts and measurements on the final state. The QCD dynamics in the soft and collinear limits, at leading order in the expansion parameter $\lambda$, are encoded in the SCET operator matrix element. At this point, the renormalization group evolution (RGE) of the hard Wilson coefficient can be used to sum logarithms of the form $\ln(\mu_{cs}/m_H) \sim \ln\lambda$ that arise as the ratio of some low soft-collinear scale $\mu_{sc} \sim \lambda m_H$ (set by the measurement in the matrix element) and the hard interaction scale $\sim m_H$.

The matrix element in \eq{dsigma} contains further logarithms of $\lambda$ due to the different scaling of soft and collinear contributions. These logarithms are part of the full logarithmic structure of the cross section, and their resummation requires the separation of soft and collinear contributions. In SCET, this separation proceeds in two steps. First, the decoupling of soft emissions happens via a field redefinition of the collinear quark and gluon fields~\cite{Bauer:2001yt}, after which $\cO_{ggH}$ takes the form 
\begin{equation} \label{eq:O_fact}
\cO_{ggH} = H\,\cO_a\, \cO_s\, \cO_b
= H\,\mathcal{B}_{n_a\perp}^\mu T\bigl[\mathcal{Y}^\dagger_{n_a} \mathcal{Y}_{n_b}\bigr]\mathcal{B}_{n_b\perp\,\mu}
\,.\end{equation}
Here, $H$ is the Higgs field, $\cO_a$ and $\cO_b$ are collinear gluon fields $(\mathcal{B}_{n\perp})$ in the forward ($n_a$) and backward ($n_b$) beam directions, and $\cO_s$ is an operator of soft lightlike Wilson lines $(\mathcal{Y}_n)$ along the $n_a$ and $n_b$ directions. After the soft-collinear decoupling, the SCET Lagrangian contains no interactions between soft and collinear fields at leading order in $\lambda$.

Second, the measurement $\widehat{\cM}$ must be separated into collinear and soft components that act independently on the soft and collinear final states. Schematically,
\begin{equation} \label{eq:M_fact}
\widehat\cM = \widehat\cM_a \times \widehat\cM_b \times \widehat\cM_s + \delta \widehat\cM
\,.\end{equation}
The operators $\widehat\cM_a$, $\widehat\cM_b$, and $\widehat\cM_s$ are obtained by restricting $\widehat\cM$ to act only on $n_a$-collinear, $n_b$-collinear, and soft fields respectively.  The total contribution from the remainder $\delta\widehat\cM$ must be power suppressed in $\lambda$ (such that to all orders it can contribute at most terms that scale like $\lambda\ln^n\lambda$).

With these two ingredients, the matrix element in \eq{dsigma} factorizes into independent soft and collinear matrix elements,
\begin{align}
B_a(\mu) &\sim \Mae{p_a}{\cO_a^\dagger\,\widehat\cM_a\, \cO_a}{p_a}(\mu)
\,,\nn\\
B_b(\mu) &\sim \Mae{p_b}{\cO_b^\dagger\,\widehat\cM_b\, \cO_b}{p_b}(\mu)
\,,\nn\\
S(\mu) &\sim \Mae{0}{\cO_s^\dagger\,\widehat\cM_s\, \cO_s}{0}(\mu)
\,.\end{align}
Here $S$ is a soft function and $B$ is a beam function~\cite{Stewart:2009yx, Stewart:2010qs, Fleming:2006cd}, which describes the collinear initial-state radiation from the parton entering the hard interaction, and can be calculated perturbatively by matching onto the standard parton distribution functions [see \eq{BgPDF}].  It follows that the cross section factorizes as well,
\begin{equation} \label{eq:dsigma_fact}
\df\sigma_{gg\to H} \sim \lvert C_{ggH}(\mu) \rvert^2\, \bigl[ B_a(\mu) \times B_b(\mu) \times S(\mu) \bigr]
\,.\end{equation}
The collinear and soft matrix elements are renormalized objects and depend on a renormalization (or separation) scale $\mu$. The RGE running between the natural collinear and soft scales then resums logarithms of the form $\ln(\mu_c/\mu_s) \sim \ln\lambda$ that are present in the soft-collinear matrix element itself.

From this discussion, we can see that resumming the logarithmic series in $\ln\lambda$ requires an explicit separation of the degrees of freedom whose associated scale depends on $\lambda$. In the effective-field theory context the resummation is then achieved by RGE methods. The complete all-order resummation requires this separation to hold to all orders in perturbation theory.  This is guaranteed for the hard-scattering factorization in \eq{dsigma}, which essentially amounts to expanding QCD in the soft-collinear limit. It also holds for the soft-collinear operator decoupling, which is independent of the measurement.  However, for the measurement function it means that the separation in \eq{M_fact} has to hold for any number of soft and collinear particles in the final state. As we will see below, this requirement is satisfied by the inclusive observables, but for the jet-based observables it provides a nontrivial constraint and only holds for $R\sim\lambda$.

In the following we discuss the soft-collinear factorization properties of our four observables and the resulting factorization formulas for the cross section. We consider the inclusive observables in \subsec{inclobservables} and the jet-based ones in \subsec{exclobservables}. In \app{RGconstraints} we discuss the RGE constraints on the factorization formulas for these observables.

\subsection{Inclusive Observables}
\label{subsec:inclobservables}

The full measurement operator $\widehat\cM$ entering \eq{dsigma} directly follows from the definitions of the observables in \eqs{inclvarTB}{inclvarET}. For example, for beam thrust its action on a given final state with a set of momenta $\{p_m\}$ is
\begin{equation}
\cM(\Tau_B) = \delta\Bigl(\Tau_B - \sum_{m} \abs{\vec{p}_{Tm}}\,e^{-\abs{y_m - Y}} \Bigr)
\,,\end{equation}
and analogously for $E_T$. Since the inclusive observables simply sum over particles, we can write them as a sum over separate contributions from $n_a$-collinear, $n_b$-collinear, and soft particles,
\begin{align}
\Tau_B &= \Tau_{Ba} + \Tau_{Bb} + \Tau_{Bs}
\,,\nn\\
E_T &= E_{Ta} + E_{Tb} + E_{Ts}
\,.\end{align}
This means there is no contribution $\delta \cM$ at leading power which mixes the different sectors,
and we can factorize the measurement function in terms of a convolution,
\begin{align}\label{eq:Mincl}
\cM(k)
&= \int\!\df k_a\, \df k_b\, \df k_s\,
\delta(k - k_a - k_b - k_s)
\nn\\ & \qquad\times
\cM_a(k_a)\,\cM_b(k_b)\,\cM_s(k_s)
\,,\end{align}
where $k$ here stands for either $\Tau_B$ or $E_T$.  In the case of $\Tau_B$ we have
\begin{align}
\cM_a(\Tau_{Ba})
&= \delta\Bigl(\Tau_{Ba} - \sum_{m\in n_a-\mathrm{coll}} \abs{\vec{p}_{Tm}}\,e^{Y-y_m} \Bigr)
\,,\nn\\
\cM_a(\Tau_{Bb})
&= \delta\Bigl(\Tau_{Bb} - \sum_{m\in n_b-\mathrm{coll}} \abs{\vec{p}_{Tm}}\,e^{y_m-Y}\Bigr)
\,,\nn\\
\cM_s(\Tau_{Bs})
&= \delta\Bigl(\Tau_{Bs} - \sum_{m\in\mathrm{soft}} \abs{\vec{p}_{Tm}}\,e^{-\abs{y_m - Y}} \Bigr)
\,.\end{align}
Note that for the collinear sectors, the contribution from the respective opposite hemisphere does not contribute, because it is power-suppressed. Similarly, for $E_T$ we have
\begin{align}
\cM_a(E_{Ta})
&= \delta\Bigl(E_{Ta} - \sum_{m\in n_a-\mathrm{coll}} \abs{\vec{p}_{Tm}} \Bigr)
\,,\nn\\
\cM_a(E_{Tb})
&= \delta\Bigl(E_{Tb} - \sum_{m\in n_b-\mathrm{coll}} \abs{\vec{p}_{Tm}} \Bigr)
\,,\nn\\
\cM_s(E_{Ts})
&= \delta\Bigl(E_{Ts} - \sum_{m\in\mathrm{soft}} \abs{\vec{p}_{Tm}} \Bigr)
\,.\end{align}

Using $n_a$ and $n_b$ we define light-cone coordinates,
\begin{equation}
p^\mu = (n_a\cdot p, n_b\cdot p, p_\perp)
\,,\quad
p^\mu = n_a\cdot p\,\frac{n_b^\mu}{2} + n_b\cdot p\,\frac{n_a^\mu}{2} + p_\perp^\mu
\,.\end{equation}
In terms of these, $n_a$-collinear particles have momentum scaling $p_a \sim m_H(\lambda^2, 1, \lambda)$, while $n_b$-collinear particles have momentum scaling $p_b \sim m_H(1, \lambda^2, \lambda)$. The soft and collinear contributions to $\Tau_B$ can be written as
\begin{align}
\Tau_{Ba} &= e^Y n_a\cdot P_a
\,,\qquad
\Tau_{Bb} = e^{-Y} n_b \cdot P_b
\,,\nn\\
\Tau_{Bs}
&= \sum_{m\in\mathrm{soft}} \min\{e^Y n_a\cdot p_m,\, e^{-Y} n_b\cdot p_m\}
\,,\end{align}
where $P_{a,b}$ is the total momentum of all $n_{a,b}$-collinear final-state particles. Since $\Tau_B$ measures the small light-cone components, soft particles that contribute to the measurement of $\Tau_B$ have ultrasoft ($us$) momentum scaling $p_{us} \sim m_H(\lambda^2, \lambda^2, \lambda^2)$. The appropriate version of SCET for this case is called \SCETI. Since $p_{a,b}^2 \sim \lambda^2 m_H^2 \gg p_{us}^2 \sim \lambda^4 m_H^2$, soft and collinear degrees of freedom in \SCETI are separated in virtuality and the RGE running in \SCETI describes evolution in invariant mass.

The factorization formula for the beam thrust distribution is~\cite{Stewart:2009yx, Berger:2010xi}
\begin{align} \label{eq:TauB_fact}
\frac{\df\sigma}{\df \Tau_B}
&= \sigma_0\, H_{gg}(m_H, \mu)  \int\!\df Y \int\!\df k_a\, \df k_b
\nn\\ &\quad \times
B_g(m_H k_a, x_a, \mu)\, B_g(m_H k_b, x_b, \mu)
\nn\\ &\quad \times
S_B^{gg}(\Tau_B - k_a - k_b, \mu)
\,,\end{align}
where
\begin{equation}\label{eq:xdef}
x_a = \frac{m_H}{\Ecm}\,e^{Y}
\,,\qquad
x_b = \frac{m_H}{\Ecm}\,e^{-Y}
\,,\end{equation}
and
\begin{equation}\label{eq:sigma0def}
\sigma_0 = \frac{\sqrt{2} G_F\, m_H^2}{576 \pi \Ecm^2}
\,.\end{equation}
The convolution between beam and soft functions in \eq{TauB_fact} is a direct consequence of the convolution of the measurement in \eq{Mincl}.

In contrast to $\Tau_B$, $E_T$ measures the perpendicular component of momentum, which means soft particles contributing to the measurement of $E_T$ have soft momentum scaling $p_s \sim m_H(\lambda, \lambda, \lambda)$. The appropriate version of SCET for this case is called \SCETII. In \SCETII, soft and collinear degrees of freedom have the same virtuality scaling, $p_{a,b}^2 \sim p_s^2 \sim \lambda^2 m_H^2$, but are still separated in rapidity. As a result, the RGE running in \SCETII describes both evolution in invariant mass as well as rapidity~\cite{Chiu:2011qc, Chiu:2012ir}.

The factorization formula for the $E_T$ distribution is~\cite{Papaefstathiou:2010bw}
\begin{align} \label{eq:ET_fact}
\frac{\df\sigma}{\df E_T}
&= \sigma_0\, H_{gg}(m_H, \mu)  \int\!\df Y \int\!\df k_a\, \df k_b\,
\nn\\ &\quad \times
B_g(m_H,k_a, x_a, \mu, \nu)\, B_g(m_H,k_b, x_b, \mu, \nu)
\nn\\ &\quad \times
S_T^{gg}\Bigl(E_T - k_a - k_b, \mu, \nu\Bigr)
\,.\end{align}
The soft and beam functions are different from those in \eq{TauB_fact} because they contain a different measurement. Beam functions with $p_T$-dependence were studied in SCET in Refs.~\cite{Mantry:2009qz, Mantry:2010mk, Becher:2010tm, Jain:2011iu, GarciaEchevarria:2011rb, Becher:2012qa, Chiu:2011qc, Chiu:2012ir}. Since the beam and soft functions in \eq{ET_fact} are renormalized and RG evolved in rapidity, they depend on another scale $\nu$ from this running.

Higher-order resummation at small $\Tau_B$ and $E_T$ can be carried out systematically using \eqs{TauB_fact}{ET_fact}. Extending the $\Tau_B$ resummation to N$^3$LL requires determining the two-loop beam and soft functions, while resuming $E_T$ to NNLL requires determining the two-loop beam and soft non-cusp anomalous dimensions.

\subsection{Exclusive Observables}
\label{subsec:exclobservables}

Vetoes on individual jets rather than an inclusive veto on the final state requires a more careful understanding of the role of the jet algorithm.
When putting a cut on $\Tauj$ or $\pTj$ in \eqs{exclvarTau}{exclvarpT}, the full measurement operator acting on the complete final state in \eq{dsigma} is
\begin{align} \label{eq:MvetoFull}
\cM^\jet(\Taucut)
&= \prod_{m\in j(R)} \theta \left( \abs{\vec{p}_{Tm}} e^{-\lvert y_m - Y\rvert} < \Tau^\cut \right)
\,, \nn \\
\cM^\jet(\pTcut) &= \prod_{m\in j(R)} \theta \left( \abs{\vec{p}_{Tm}} < \pTcut \right)
\,,\end{align}
where we consider jets $j(R)$ defined by the $\kt$ class of algorithms. Note that integrating $\Tauj$ or $\pTj$ up to a cut turns the maximum condition in their definitions in \eqs{exclvarTau}{exclvarpT} into the simple product of $\theta$ functions in \eq{MvetoFull}. Hence, for the exclusive variables it is more convenient to consider the integrated cross section with a cut rather than the differential spectrum as in the inclusive variables.

We now want to separate the measurement function into components that act independently on soft and collinear final states. That is, we want to express the full jet veto in \eq{MvetoFull} as
\begin{align}\label{eq:Mjet_fact}
\cM^\jet(k^\cut)
&= \cM^\jet_a(k^\cut)\, \cM^\jet_b(k^\cut)\, \cM^\jet_s(k^\cut)
\nn\\ &\quad
+ \delta \cM^\jet(k^\cut)
 \,,\end{align}
where here and in the remainder of this section, $k^\cut$ stands for either $\Taucut$ or $\pTcut$. The functions $\cM^\jet_{i}(k^\cut)$ for $i=a,b,s$ are defined by the full measurement applied to $n_a$-collinear, $n_b$-collinear, and soft particles, respectively, which also defines the remainder $\delta \cM^\jet(k^\cut)$. For this separation to be meaningful, the remainder should be power-suppressed, which requires that the full measurement $\cM^\jet$ does not mix constraints on collinear and soft particles. This is a nontrivial condition, since the veto on any individual jet is not allowed to mix constraints between sectors. If a jet has a collinear component $p_c$ and a soft component $p_s$, then the veto condition, e.g. for $\pTcut$, is
\begin{equation} \label{eq:pTcsVeto}
\lvert \vec{p}_{Tc} + \vec{p}_{Ts} \rvert < \pTcut
\,.\end{equation}
This prevents $\cM^\jet$ from factorizing into separate soft and collinear components and gives a contribution to $\delta \cM^\jet$. Therefore, to preserve factorization this scenario should only happen with a power-suppressed rate, such that at leading power each jet contains either only soft or only collinear final states. In that case, we can perform the veto separately on jets in each sector, i.e., we can write the product over all jets in \eq{MvetoFull} as products over soft and collinear jets as in \eq{Mjet_fact}.

We shall show that for $R\sim \lambda$, $\delta \cM^\jet$ indeed gives a power-suppressed contribution to the rate. In this limit, the following factorization formula holds for the $\Tauj$ veto:
\begin{align}\label{eq:VetoTau_Fact}
\sigma(\Taucut)
&= \sigma_0 H_{gg} (m_H, \mu) \int dY \, B^\jet_g \left( m_H\Taucut, x_a, \mu \right)
\nn \\ & \quad \times
B^\jet_g \left( m_H\Taucut, x_b,\mu \right) S^\jet_{gg}(\Taucut, \mu)
\,,\end{align}
where $x_a,x_b$ and $\sigma_0$ are defined in \eqs{xdef}{sigma0def}. In the same limit, an analogous factorization formula holds for the $\pTj$ veto:
\begin{align}\label{eq:VetopT_Fact}
\sigma(\pTcut)
&= \sigma_0 H_{gg} (m_H, \mu) \int dY \, B^\jet_g \left(m_H,\pTcut, x_a, \mu,\nu \right)
\nn \\ & \quad \times
B^\jet_g \left( m_H,\pTcut, x_b,\mu,\nu \right) S^\jet_{gg}(\pTcut, \mu,\nu)
\,.\end{align}
An equivalent form of this factorization formula was derived in Ref.~\cite{Becher:2012qa}, and the NLL resummation for $\pTcut$ was performed in Refs.~\cite{Banfi:2012yh, Becher:2012qa}.

In Ref.~\cite{Banfi:2012yh} jet-algorithm dependent effects were calculated at fixed $\ord{\as^2}$. These results were interpreted in Ref.~\cite{Becher:2012qa} in terms of a two-loop anomalous dimension and used to extend the resummation based on \eq{VetopT_Fact} to NNLL, working in the limit $\lambda \ll R \sim 1$ to avoid $\ln R$ clustering logarithms. However, we will show explicitly in the following that for $R\sim 1$ \eqs{VetoTau_Fact}{VetopT_Fact} do not reproduce the all-order structure of QCD beyond NLL. We will also see that for $R\sim\lambda$, where the factorization formula holds, only some parts of the $\ord{\alpha_s^2}$ contributions from Ref.~\cite{Banfi:2012yh} are correctly interpreted in terms of anomalous dimensions and used in conjunction with \eq{VetopT_Fact}.

Since $\cM^\jet$ in \eq{Mjet_fact} is a simple product, and not a convolution as for the inclusive variables in \eq{Mincl}, the factorized cross sections in \eqs{VetoTau_Fact}{VetopT_Fact} now contain a product of beam and soft functions rather than a convolution. Also, each function explicitly depends on the jet algorithm used, in addition to the jet-veto variable itself, and includes $R$-dependent clustering effects. Note that the $\ord{\alpha_s}$ results do not yet depend on the effects of the jet algorithm. Nevertheless, the resummed cross sections are different for the inclusive and exclusive observables starting at NLL because of the different structures of their factorization theorems. This reflects the fact that constraining the sum of emissions provides a very different phase-space constraint than constraining each individual emission for more than one emission.

To understand the role of the jet algorithm in vetoing on individual jets and how it impacts the logarithmic series, it is useful to express the measurement function in the form
\begin{align}
\cM^\jet &=
( \cM_a + \Delta \cM_a^\jet)\, (\cM_b + \Delta \cM_b^\jet)\, ( \cM_s + \Delta \cM_s^\jet )
\nn\\ & \quad
+ \delta \cM^\jet
\,,\end{align}
where $\Delta \cM_i^\jet$ is defined to contain the jet-algorithm dependence within each of the collinear and soft sectors,%
\begin{equation}
\cM^\jet_i(k^\cut) = \cM_i (k^\cut) + \Delta \cM_i^\jet (k^\cut)
\,,\end{equation}
for $i=a,b,s$.  The definition of $\Delta \cM_i^\jet$ is subtle, since it depends on what we define the corrections due to clustering \emph{relative} to; namely it depends on the precise choice of $\cM_{i}$, which is independent of $R$ and the jet algorithm. To study the effect of clustering we choose $\cM_i$ to be the inclusive $\Tau_B$ or $E_T$ measurement,
\begin{equation}
\cM_i(k^\cut) = \theta\biggl(\sum_{m \in i^{th} - {\rm sector}} k_m < k^\cut\biggr)
\,,\end{equation}
where $k_m$ is the $\Tau_B$ or $E_T$ contribution from each particle.

At $\ord{\as}$ there are no jet-algorithm effects, since there is only a single, either soft or collinear, final-state particle. This means that $\delta \cM^{(1)}$ and $\Delta \cM_i^{(1)}$ are zero, and $\cM^{\jet \, (1)}$ reduces to a sum over terms with one nontrivial constraint in each sector,
\begin{equation}
\cM^{\jet \, (1)}(k^\cut) = \sum_{i=a,b,s}  \theta(k_i < k^\cut)
\,.\end{equation}
This gives the same $\ord{\as}$ contribution as the integral over the $k = \Tau$ or $E_T$ distribution with $k<k^\cut$.

\begin{figure}[t]
\centering
\includegraphics[scale=0.33]{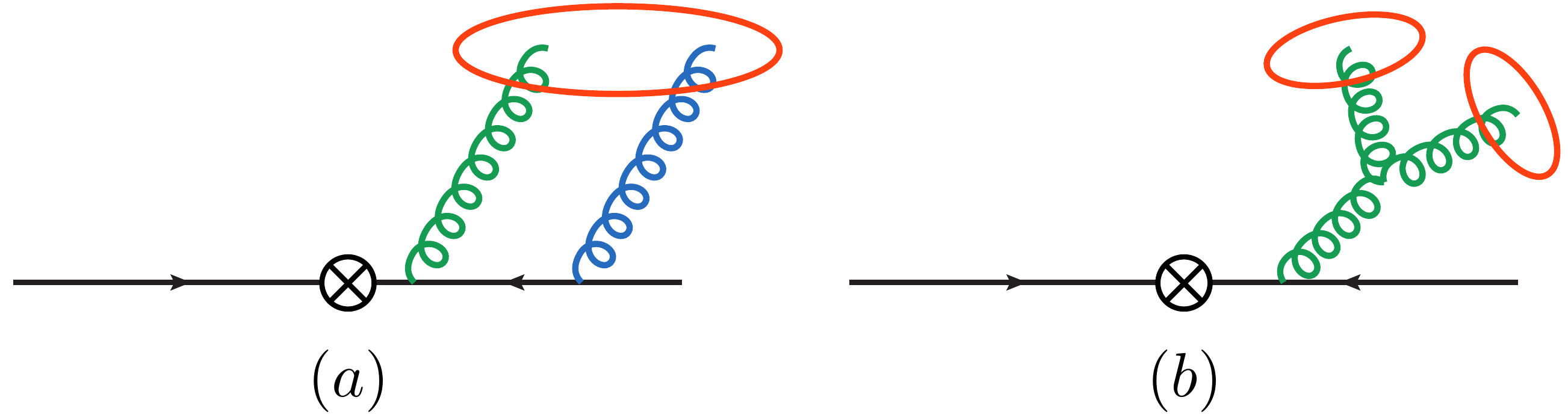}
\vspace{-0.5ex}
\caption{(a) The contribution to soft-collinear mixing at $\as^2$ from the independent emission of a collinear and soft gluon clustered into a single jet, relevant when $R\gg \lambda$. (b) Clustering corrections in the beam and soft functions from correlated emissions, relevant when $R\sim \lambda$. }
\label{fig:clustering}
\end{figure}

Starting at $\ord{\as^2}$ the role of the jet algorithm must be understood. When $R\gg \lambda$, soft-collinear mixing effects are important. They give a correction to the cross section of the form
\begin{align} \label{eq:dsigSC}
\delta \sigma^\jet_{SC}
&\sim \lvert C_{ggH}(\mu) \rvert^2
\nn\\ & \quad\times
\Mae{p_ap_b}{\cO_{ggH}(\mu)^{\dagger} \delta\widehat{\cM}^\jet \, \cO_{ggH}(\mu) }{p_ap_b}
\,,\end{align}
where the measurement operator $\delta \widehat{\cM}^\jet$ is defined by \eq{Mjet_fact} as the difference between the full measurement and its restriction to the soft and collinear sectors.  At $\ord{\as^2}$, the soft-collinear mixing contribution arises from the independent emission of a collinear and soft particle as shown in \fig{clustering}(a). It is given by the measurement function
\begin{equation} \label{eq:deltaM2}
\delta \cM^{\jet\,(2)} (k^\cut)=  \delta \cM^{\jet\,(2)}_{as} (k^\cut)+\delta \cM^{{\rm jet}\,(2)}_{bs}(k^\cut)
\,,\end{equation}
where the two terms correspond to a soft particle clustering with either a $n_a$-collinear or $n_b$-collinear particle.
For the $\kt$-class algorithms, which cluster particles with $\Delta R < R$,
\begin{align}\label{eq:deltaM2as}
\delta \cM^{{\rm jet}\,(2)}_{as} (k^\cut) &=
\theta(\Delta R_{as} <R)\, \bigl[ \theta(k_{\rm jet}< k^\cut)
\nn\\ &\quad
- \theta(k_{a}<k^\cut) \theta(k_{s}<k^\cut)  \bigr]
\,,\end{align}
where $k_\jet$ is the value of the jet observable after combining the soft and collinear particles. Since $\delta \cM^\jet$ is defined as the correction to the factorized measurement, the second term subtracts the corresponding $\ord{\alpha_s^2}$ contribution from  $\cM_a^{\jet(1)} \cM_s^{\jet (1)}$.

In \sec{scmixing}, we calculate the (leading) contribution from \eq{deltaM2as} to the cross section explicitly, and we will see that it has the form $\sim\as^2 R^2 \ln(m_H/k^\cut)$ for both $\Tauj$ and $\pTj$ vetoes. There are also factorized clustering corrections of the same form in $\Delta \cM_i^\jet$ (for $i=a,b,s$) from two independent emissions within each sector, and which are part of the two-loop soft and beam functions. The total QCD contribution from independent emissions thus arises as the sum of the factorized contributions and the soft-collinear mixing contribution. This shows explicitly that the factorized cross sections in \eqs{VetoTau_Fact}{VetopT_Fact} do not reproduce the full NNLL structure of QCD for general $R\sim 1$, since they do not contain the soft-collinear mixing contributions.

When $R\sim \lambda$, clustering corrections from independent emissions, including the soft-collinear mixing term $\delta \cM^\jet$, can be regarded as power corrections. In this limit a different type of clustering corrections in $\Delta \cM_i^\jet$ become important, arising from correlated emissions within the beam and soft functions, as shown in \fig{clustering}(b). At $\ord{\as^2}$ they are described by the measurement function
\begin{align} \label{eq:DeltaM2}
&\Delta \cM_i^{\jet \, (2)} (k^\cut)
\nn\\ & \qquad
= \biggl\{\theta(\Delta R_{12} < R) \theta(k_\jet < k^\cut)
\nn \\ & \qquad\quad
+ \theta(\Delta R_{12} > R) \theta(k_1 < k^\cut) \theta(k_2 < k^\cut) \biggr\}
\nn \\
& \qquad\quad
- \theta (k_1 + k_2 < k^\cut)
\,,\end{align}
where particles $1$ and $2$ both belong to sector $i$. In \sec{clusteringlogs} we calculate the contributions from \eq{DeltaM2}, and we will see that they have the form $\as^2 \ln R \ln(m_H/k^\cut)$. We also discuss the higher-order structure of clustering logarithms and argue that at $\ord{\alpha_s^n}$ terms of the form $\as^n \ln^{n-1} R \ln(m_H/k^\cut)$ contribute in the exponent of the cross section. For $R\sim \lambda$ these terms are NLL. Since the coefficients of the clustering logarithms at each order contain a genuinely new algorithm-dependent contribution, which will generically be unrelated to lower orders, they cannot be resummed with present methods~\cite{Kelley:2012zs}.

\subsubsection{Rapidity Cutoffs on Jets}

In an experiment, the physical limitations of the detector impose a rapidity cutoff on measured jets.  The parton luminosities naturally suppress forward, high-$p_T$ jets, but in experimental analyses even moderate rapidity cutoffs (e.g., a cutoff of $y_\cut = 2.5$) are used.  Monte Carlo studies of the dependence on the cutoff with a $p_T$ veto on jets find a negligible effect for $y_\cut \sim 4.5$ but an $\ord{10\%}$ effect for $y_\cut \sim 2.5$ in the range of typical $\pTcut$ values~\cite{Berger:2010xi, Banfi:2012yh}.

The rapidity cutoff $y_\cut$ regulates the rapidity divergences present in the soft and beam functions for the $\pTj$ veto.  In the bare functions, this effectively converts rapidity divergences into factors of $y_\cut$.  At $\ord{\as}$, the RG structure of the hard, beam, and soft functions implies that these divergences have no effect on the fixed-order logarithms.  The first order that the effect of the rapidity cutoff on the resummation can be observed is at $\ord{\as^2}$, and, in principle, the cutoff could affect the $\ord{\as^2 \ln^2\!\lambda}$ and $\ord{\as^2 \ln \lambda}$ terms.  Although we do not consider these effects here, it would be interesting to study their impact on the logarithmic structure and resummation in more detail.

\section{Soft-Collinear Mixing}
\label{sec:scmixing}

The jet algorithm used to define $\Tauj$ and $\pTj$ gives rise to jets that can contain both soft\footnote{Our discussion in this section is mostly insensitive to whether the soft radiation is described by ultrasoft modes (a $\Tauj$ veto, \SCETI) or soft modes (a $\pTj$ veto, \SCETII), but we will point out when subtleties arise.} and collinear particles, and the collinear beam radiation can give rise to multiple jets.  Placing a veto on such jets mixes the phase space constraints in each sector through conditions of the form%
\begin{equation} \label{eq:Cutmix}
\Tau_c + \Tau_s  \leq \Taucut
\,, \qquad
\lvert \vec{p}_{Tc} + \vec{p}_{Ts} \rvert \leq \pTcut
\,,\end{equation}
and this mixing contributes to $\delta \cM^\jet$ in \eq{Mjet_fact}.  In this section we show that these soft-collinear mixing contributions scale like $R^2$, and so $R$ must scale as $\lambda$ for these effects to be power suppressed.

Clustering jet algorithms build jets by merging particles in the final state according to a distance metric $\rho_{ij}$ between particles and a metric $\rho_i$ for single particles.  In each clustering step, the minimum metric determines the action of the algorithm. If the minimum is a pairwise metric then that pair is merged into a new particle, and if the minimum is a single particle then that particle is promoted to a jet.  For the anti-$\kt$ algorithm~\cite{Cacciari:2008gp}, these metrics are
\begin{equation} \label{eq:akTmetrics}
\rho_{ij} = \min \bigl( p_{Ti}^{-1}, p_{Tj}^{-1} \bigr) \frac{\Delta R_{ij}}{R}
\,, \qquad
\rho_i = p_{Ti}^{-1}
\,.\end{equation}
The algorithm will cluster particles with separation $\Delta R_{ij} < R$ together.%
\footnote{In some cases, a clustering sequence can pull two particles farther than $R$ away from each other.  Such configurations tend to be uncommon.}

Phase-space constraints from jet algorithms have been studied in SCET in Refs.~\cite{Ellis:2010rwa, Ellis:2009wj, Cheung:2009sg, Jouttenus:2009ns}. By applying the canonical scaling of soft and collinear particles to the anti-$\kt$ metrics in \eq{akTmetrics} we can determine the \emph{typical} behavior of the algorithm. For example, for the case of a veto on the jet $p_T$, the metrics scale as
\begin{align}\label{eq:metricScaling}
\rho_{cc} \sim \rho_{ss} \sim \frac{\lambda^{-1}}{R}
\,, \quad
\rho_{sc} \sim  \frac{\lambda^{-1}}{R} \ln\frac{1}{\lambda}
\,, \quad
\rho_{c}\sim\rho_{s} \sim \lambda^{-1}
\,.\end{align}
This naive power counting suggests that clustering of soft and collinear particles will typically not occur, since soft particles are at central rapidities and collinear particles are at large rapidities and as a result $\rho_{sc} \gg \{\rho_{cc},\rho_{ss}, \rho_c,\rho_s\}$ (for $R \sim 1$ or smaller).  The ordering $\rho_{sc} > \rho_{ss}, \rho_{cc}$ implies that collinear and soft particles will separately cluster among themselves before clustering with each other. The ordering $\rho_{sc} > \rho_{c}, \rho_s$ implies that (groups of) collinear and soft particles will be promoted to jets before any soft-collinear clustering between them can take place.  A similar argument applies to measuring $\Tauj$.

Power counting along these lines, counting $\lambda \ll R \ll \ln(1/\lambda)$, was used in Ref.~\cite{Becher:2012qa} to argue that, at leading power, soft and collinear particles do not occupy the same jet. However, the canonical scaling in \eq{metricScaling} is not sufficient to demonstrate that soft collinear clustering gives only a power-suppressed contribution to the cross section~\cite{Walsh:2011fz}. In particular, there is a collinear enhancement in soft emissions along the beam directions (i.e., the direction of the soft Wilson lines).  This is straightforward to see from the matrix element for eikonal emission,
\begin{equation}
\cA_{\rm eikonal} \sim \frac{\df p_T}{p_T} \, \df y \, \df\phi
\,.\end{equation}
The matrix element is flat in rapidity, meaning soft particles will populate jets at a forward rapidities at an equal rate to jets at central rapidities.  For a jet $j$ in a given direction, the rate for a soft particle to be radiated within the cone of the jet scales as
\begin{equation} \label{eq:mixingscaling}
\int_{-\infty}^{\infty} \df y_s \int_0^\pi \frac{\df \phi_s}{\pi} \, \theta(\Delta R_{js}<R) \sim R^2 \,.
\end{equation}
If $R \sim \lambda$, this rate is power suppressed and can be neglected.  We will see this scaling confirmed in the explicit calculations of soft-collinear mixing below.  Note that the counting $R\sim1 \ll \ln(1/\lambda)$ is insufficient to make this contribution power suppressed.

\begin{figure}[t]
\centering
\includegraphics[scale=0.28]{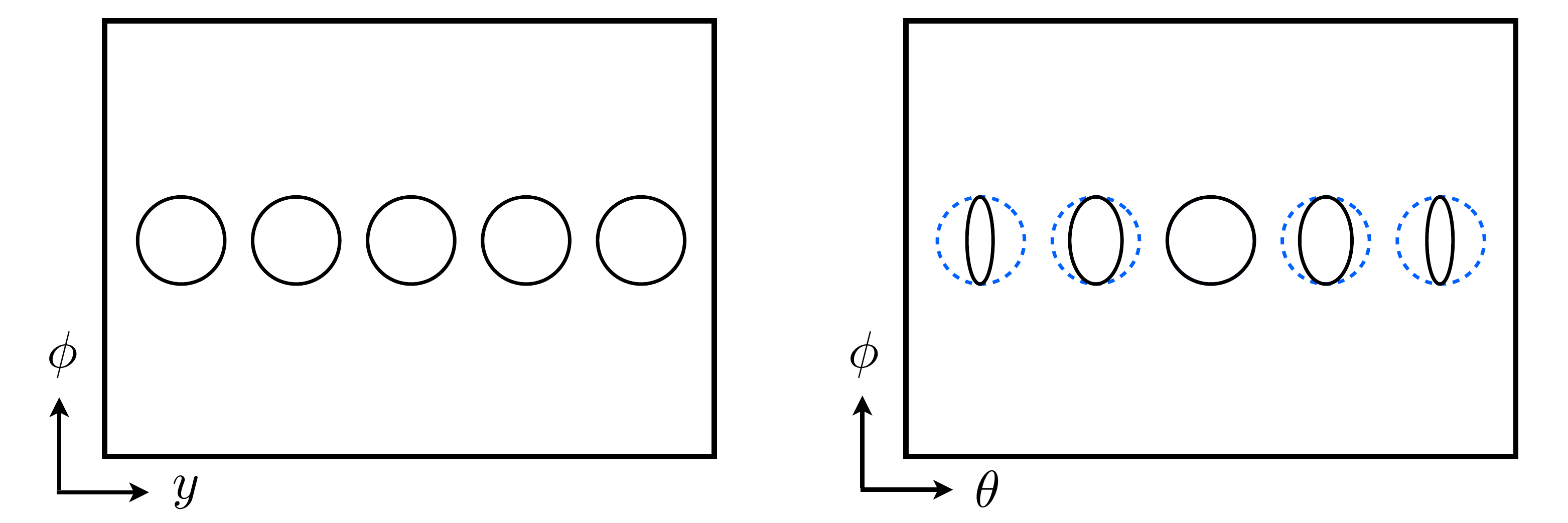}
\vspace{-0.5ex}
\caption{The shape of jets in two spaces: $y-\phi$ (left), the geometry where jets are found, and $\theta-\phi$ (right), which is the physical geometry.  As jets become forward, their size in $\theta$ shrinks.  In the right figure, the dashed blue line shows the approximate shape of jets if $\Delta R_{ij}$ is replaced with the physical angle $\theta_{ij}$ between particles.}
\label{fig:JetSizes}
\end{figure}

The geometry of jets at a hadron collider changes significantly in the forward region.  For central rapidities, the clustering condition $\Delta R_{ij} < R$ is approximated by $\theta_{ij} < R$, where $\theta$ is the physical angle between $i$ and $j$.  At forward rapidities, the jet subtends a smaller angle $R_{\theta}$ with respect to the beam axis. The size of the jets in $\phi$ remains unchanged, so jets become elongated at forward rapidities.  A comparison between jets in $y-\phi$ space and $\theta-\phi$ space is shown in \fig{JetSizes}.
If a jet is at rapidity $y_j$ with a corresponding angle to the beam $\theta_j$, then
\begin{equation}
R_{\theta} \approx 2\theta_j \sinh R
\,.\end{equation}
At the characteristic angle $\theta_c \sim \lambda$ of collinear emissions, the size of the jet scales approximately as $\lambda R$. This means that collinear initial-state radiation from the incoming hard partons can create multiple jets of size $R$ in the final state, since the size of the jet is smaller than (potentially parametrically smaller than) the total size of the collinear beam sector, which is an angle of order $\lambda$ around the beam direction. This is important because if each collinear sector created a single jet (around the beam), a veto on jets containing both soft and collinear particles would require $\delta \cM^\jet$ to only depend on the total collinear momentum and the measurement could therefore be expressed in a factorizable way. However this is not the case here. From this analysis, we see that jets can contain both soft and collinear particles.  Vetoing on such jets requires a constraint on the combined soft and collinear contribution to the veto variable, as in \eq{Cutmix}. Therefore, soft-collinear factorization will be inhibited by these jets for $R\sim 1$.

We can calculate the correction to the cross section from soft-collinear mixing, $\delta \sigma^\jet_{SC}$ in \eq{dsigSC}, for both jet-based veto observables at $\ord{\as^2}$.  The mixing correction is given by the measurement function $\delta \cM^{{\rm jet} \,(2)}$ in \eqs{deltaM2}{deltaM2as} inserted into matrix elements of $\cO_{ggH}$ with one soft and one collinear emission. Because of the soft-collinear decoupling this is simply the product of the $\ord{\as}$ matrix elements for one soft and one collinear emission.

The form of the soft-collinear mixing terms are the same as the $\ord{\as^2}$ corrections when two independent emissions within either the soft or collinear sectors cluster (which is contained in $\Delta\cM_i^\jet$). Therefore, the total correction from clustering of two independent emissions to the cross section is given by
\begin{equation} \label{eq:Deltasigindep}
\Delta \sigma^{\rm indep\,(2)} = \Delta \sigma^{\rm indep\,(2)}_{SS}+\Delta \sigma^{\rm indep\,(2)}_{CC}+\delta \sigma^{\jet\,(2)}_{SC}
\,,\end{equation}
where $SS$, $CC$, and $SC$ denote the independent emission contributions from the soft sector, collinear sector, and from soft-collinear mixing, respectively. In addition to the soft-collinear mixing term, we also calculate the correction from the soft sector, $\Delta \sigma^{\rm indep}_{SS}$.%
\footnote{Determining the $\ord{\as^2}$ beam function matrix elements is significantly more involved, and is not required to demonstrate the presence of soft-collinear mixing. It is therefore beyond the scope of this work.} The calculations are given in \app{scmixingcalc}, where we work to leading power in $R$, $\ord{R^2}$, and drop corrections of $\ord{R^4}$.

The $\Tauj$ veto is easier to calculate, as dimensional regularization fully handles all divergences.  The soft-collinear mixing, $\delta \sigma^\jet_{SC}$, is the sum of contributions from each beam, $\delta \sigma^\jet_{SC_a} + \delta \sigma^\jet_{SC_b}$ and is given in \eq{DeltSigCStau}.  The two contributions $\delta \sigma^\jet_{SC_a}$ and $\delta \sigma^\jet_{SC_b}$ are identical, and their sum is
\be \label{eq:SigmaSCTau}
\delta \sigma^{\rm jet \, (2)}_{SC}(\Taucut)
= \sigma_{\rm LO} \frac{1}{\e} \Bigl(\frac{\as C_A}{\pi}\Bigr)^2 \biggl(\frac{\mu^2}{m_H \Taucut}\biggr)^{2\e} \frac{\pi^2}{6} R^2 \,.
\ee
Here $\sigma_{\rm LO}$ is the leading-order cross section, which is given in \eq{sigmaLO}. We can see explicitly that the soft-collinear mixing contains a single logarithm and scales as $R^2$, as expected from the naive scaling calculation in \eq{mixingscaling}.  The soft independent emission contribution, see \eq{DeltaSigSStau}, is
\begin{equation}
\Delta \sigma_{SS}^{\rm indep \, (2)} (\Taucut)
= - \sigma_{\rm LO}\frac{1}{\e} \Bigl(\frac{\as C_A}{\pi}\Bigr)^2
\biggl(\frac{\mu^2}{{\Taucut}^2}\biggr)^{2 \e} \frac{\pi^2}{12}R^2
\,.\end{equation}
The remaining contribution at this order is from the independent emission of two collinear particles, $\Delta \sigma_{CC}^{\rm indep}$, whose form can be constrained by two facts. First, since the effective theory reproduces the IR structure of QCD, the collinear contribution must cancel the $1/\e$ dependence in $\Delta \sigma_{SS}^{\rm indep}$ and $\delta \sigma_{SC}^{\jet}$. Second, the scale dependence of the collinear matrix element can be determined by power counting. It follows that the $\Delta \sigma_{CC}^{\rm indep}$ contribution must have the form
\begin{align}
&\Delta \sigma_{CC}^{\rm indep \, (2)} (\Taucut)
= -\sigma_{\rm LO}\frac{1}{\e} \Bigl(\frac{\as C_A}{\pi}\Bigr)^2\!
\biggl(\frac{\mu^2}{ m_H \, \Taucut}\biggr)^{2 \e} \frac{\pi^2}{12}R^2
\,.\end{align}
Combining all three contributions, the leading $R$-dependent clustering effect from independent emissions with a cut on $\Tauj$ is given by
\begin{equation}\label{eq:SigmaTauMix}
\sigma^{\rm indep \, (2)}(\Taucut)= - \sigma_{\rm LO} \Bigl(\frac{\as C_A}{\pi}\Bigr)^2  \frac{\pi^2}{6}  R^2\ln \frac{m_H}{\Taucut}
\,.\end{equation}

For the $\pTj$ veto, the total $\ord{\as^2}$ clustering effect from independent emissions in QCD, which includes the soft-collinear contributions, was calculated in Ref.~\cite{Banfi:2012yh} and found to be
\begin{equation}\label{eq:SigmaMix}
\Delta \sigma^{\rm indep \, (2)}(\pTcut)
= -\sigma_{\rm LO} \Bigl(\frac{\as C_A}{\pi}\Bigr)^2  \frac{\pi^2}{3} R^2\, \ln \frac{m_H}{\pTcut}
\,.\end{equation}
This result will serve as a partial cross check on our results. To calculate the soft-collinear mixing contribution for the $\pTj$ veto, we have to regulate rapidity divergences.  We use the analytic regulator~\cite{Becher:2011dz} in this case, for which $\nu$ plays the same role as $\mu$ in dimensional regularization, and $\alpha$ the role of $\e$.  The soft-collinear mixing contributions are different for $\delta \sigma^\jet_{SC_a}$ and $\delta \sigma^\jet_{SC_b}$, because of the asymmetry in the regulator between the two collinear sectors. We find
\begin{align} \label{eq:SigmaSCpT}
\delta \sigma^{\rm jet \, (2)}_{SC} (\pTcut)
&= \delta \sigma^{\rm jet \, (2)}_{SC_a} + \delta \sigma^{\rm jet \, (2)}_{SC_b}
\nn\\*
&=  - \sigma_{\rm LO} \frac{1}{\alpha} \Bigl(\frac{\as C_A}{\pi}\Bigr)^2\,\frac{\pi^2}{6}R^2
\nn\\* & \quad\times
\biggl[   \left(\frac{\nu \, m_H}{(\pTcut)^2}\right)^{2 \alpha} -  \Bigl(\frac{\nu}{m_H }\Bigr)^{2 \alpha} \biggr]
\nn \\
&= -\sigma_{\rm LO} \Bigl(\frac{\as C_A}{\pi}\Bigr)^2 \frac{2\pi^2}{3} R^2\,\ln \frac{m_H}{\pTcut}
\,.\end{align}
As for the $\Tauj$ veto, the soft-collinear mixing for the $\pTj$ veto contains a single logarithm and scales as $R^2$. The soft independent emission contribution for $\pTj$ is scaleless and thus vanishes
\begin{equation}
\Delta \sigma_{SS}^{\rm indep\,(2)} (\pTcut) = 0
\,.\end{equation}
Hence, to reproduce the full independent emission result in \eq{SigmaMix}, the collinear contribution must be
\begin{equation}
\Delta \sigma_{CC}^{\rm indep  \, (2)}(\pTcut)
 = \sigma_{\rm LO} \Bigl(\frac{\as C_A}{\pi}\Bigr)^2\, \frac{\pi^2}{3} R^2\,\ln \frac{m_H}{\pTcut} 
\,.\end{equation}

The soft-collinear mixing contributions in \eqs{SigmaSCTau}{SigmaSCpT} are not isolated to $\ord{\as^2}$.  At higher orders, additional emissions can generate a tower of Sudakov double logarithms on top of these results and also generate higher-order mixing effects.  Accurate uncertainty estimates require a better understanding of the soft-collinear mixing terms at all orders.

\section{Clustering Logarithms}
\label{sec:clusteringlogs}

We now turn to the form of clustering logarithms in the cross section. When $R\sim\lambda$, the cross section $\sigma (k^\cut)$ satisfies the factorization formulas in \eqs{VetoTau_Fact}{VetopT_Fact}. Clustering of final state soft and collinear particles can give rise to logarithms of $R$ that become important when $R \ll 1$.  These logarithms arise as the remnant of a collinear divergence between particles in the jet, and are associated with connected webs (c-webs) in the matrix element~\cite{Gatheral:1983cz, Frenkel:1984pz}.

To see this, consider a two-parton final state with a $\pTj$ veto.  The measurement function can be written as a constraint on each particle's $p_T$ plus a correction factor for when they are clustered into a jet,
\begin{equation} \label{eq:clusMveto}
\cM^{{\rm jet} \, (2)} = \cM_{\rm veto}^{(2)} + \Delta \cM^{{\rm jet} \, (2)}_{\rm veto} \,,
\end{equation}
where
\begin{align}
\cM^{(2)}_{\rm veto} &= \theta(p_{T1} < \pTcut)\, \theta(p_{T2} < \pTcut)
\,,\nn\\
\Delta \cM^{{\rm jet} \, (2)}_{\rm veto} &= \theta(\Delta R_{12} < R) \Bigl[ \theta(\lvert \vec{p}_{T1} + \vec{p}_{T2} \rvert < \pTcut)
\nn \\ & \quad
- \theta(p_{T1} < \pTcut)\, \theta(p_{T2} < \pTcut) \Bigr]
\,.\end{align}
Each individual measurement, $\cM^{(2)}_{\rm veto}$ and $\Delta \cM^{{\rm jet} \, (2)}_{\rm veto}$, is separately IR unsafe if the matrix element has a collinear singularity between the two particles.  Since the total measurement, $\cM^{{\rm jet} \, (2)}$, is IR safe, the divergence cancels.  The remnant is logarithmic sensitivity to $R$, schematically
\begin{align}
&\cM^{(2)}_{\rm veto} \sim \frac{1}{\e} \,, \qquad \Delta \cM^{{\rm jet} \, (2)}_{\rm veto} \sim -\frac{1}{\e} R^{-2\e} \,, \nn \\
&\cM^{(2)}_{\rm veto} + \Delta \cM^{{\rm jet} \, (2)}_{\rm veto} \sim \ln R \,.
\end{align}
This structure persists for more particles in the final state.  In general, if there are $n$ final-state particles, there are at most $n-1$ collinear singularities between them, each of which leads to a factor of $\ln R$ from clustering effects.  We will show that the general form of the leading clustering logarithm at $\ord{\alpha_s^n}$ is
\begin{equation} \label{eq:cluslogform}
\sigma_{\rm LO} \left(\frac{\as}{\pi}\right)^n \, C_{p_T}^{(n)} (\ln R) \, \ln \frac{m_H}{\pTcut}
 \,,\end{equation}
where $C_{p_T}^{(n)}(\ln R)$ contains at most $n-1$ logarithms of $R$ and the same form holds for $\Taucut$.

\subsection{\boldmath $\ord{\alpha_s^2}$ Clustering Logarithms}
\label{subsec:cluslogsalpha2}

Although the division in \eq{clusMveto} helps determine the order of the clustering logarithms, it is not well suited to define the clustering correction.  Instead, we use the division in \subsec{exclobservables}, defining the clustering logarithms relative to the inclusive measurement.

At $\ord{\as^2}$, the clustering logarithms for the $p_T$ veto have been calculated in Ref.~\cite{Banfi:2012yh}.  We perform the soft function clustering calculation for both $\pTj$ and $\Tauj$ in \app{clusteringlogcalc}, since it is instructive to see the RG structure explicitly.  The bare soft function corrections for $\Tauj$ are given in \eq{softfunctionclusTau} and for $\pTj$ in \eq{softfunctioncluspT}, where for the latter we use the rapidity regulator~\cite{Chiu:2011qc, Chiu:2012ir}. The bare corrections are UV-divergent and give a contribution to the two-loop soft anomalous dimensions of
\begin{align}
\Delta \gamma_S^{(2)} (\Taucut,\mu) &= \left(\frac{\as}{\pi}\right)^2 C_{\Tau}^{(2)} (\ln R)
\,, \nn \\
\Delta \gamma_S^{\nu(2)} (\pTcut,\nu) &= \left(\frac{\as}{\pi}\right)^2 C_{p_T}^{(2)} (\ln R)
\,,
\end{align}
where the coefficients are given by
\begin{align} \label{eq:C2coeffs}
C_{\Tau}^{(2)} (\ln R)
&= C_A^2 \Bigl( \frac{131-12\pi^2 - 132\ln 2}{18} \ln R - 0.936 \Bigr)
\nn\\ &\quad
+ C_A T_R n_f \Bigl( -\frac{23 - 24 \ln 2}{9}\, \ln R + 0.748 \Bigr)
\,, \nn \\
C_{p_T}^{(2)} (\ln R)
&= C_A^2 \Bigl( \frac{131-12\pi^2 - 132\ln 2}{18} \ln R - 1.12 \Bigr)
\nn\\ & \quad
+ C_A T_R n_f \Bigl( -\frac{ 23 - 24 \ln 2}{9}\ln R + 0.764 \Bigr)
\,.\end{align}
For the $p_T$ veto, the contribution to the cross section stemming from this coefficient agrees with the results in Ref.~\cite{Banfi:2012yh} for the $\ln R$ term. The constant terms depend the observable the clustering effect is defined relative to. We use the inclusive $E_T$ measurement, as in \eq{DeltaM2}, for this purpose, whereas Ref.~\cite{Banfi:2012yh} uses the total $p_T$, which for two particles is $p_T = \abs{\vec{p}_{T1} + \vec{p}_{T2}}$.  Either choice is possible and we have checked that we reproduce the constant terms in Ref.~\cite{Banfi:2012yh} when alternatively using their definition of the clustering correction.

The soft anomalous dimension from these clustering effects must be canceled by the anomalous dimensions of the beam functions, since the total $\cO_{ggH}$ matrix element has only UV divergences that match the hard function, and which are unrelated to clustering.  It is worthwhile to note that since no collinear singularities exist between soft and collinear particles, the cancellation of the divergences in the soft function from clustering must come entirely from the beam functions. It cannot come from a soft-collinear mixing term, which is power suppressed when $R\sim\lambda$.  The cancellation of the beam and soft anomalous dimensions from clustering leaves a fixed-order contribution that has a logarithm of the ratio of the beam and soft scales.  Note that the scale accompanying $\mu$ or $\nu$ in the fixed-order beam and soft functions is fixed by power counting. For the $p_T$ veto, the beam $\nu$ scale is $m_H$ and the soft $\nu$ scale is $\pTcut$, while for the $\Tau$ veto, the beam scale is $\sqrt{m_H \Taucut}$ and the soft scale is $\Taucut$. Hence, the clustering effect at $\ord{\as^2}$ in the cross section is
\begin{align}
\Delta \sigma^{(2)} (\Taucut)
&= \sigma_{\rm LO} \left( \frac{\as}{\pi} \right)^2 \ln \frac{m_H}{\Taucut} \, \frac12 C_{\Tau}^{(2)} (\ln R)
\,, \nn \\
\Delta \sigma^{(2)} (\pTcut)
&= \sigma_{\rm LO} \left( \frac{\as}{\pi} \right)^2 \ln \frac{m_H}{\pTcut} \, C_{p_T}^{(2)} (\ln R)
\,.\end{align}
This form matches that in \eq{cluslogform}.  Because these terms are connected with the anomalous dimension, it is possible to resum the logarithm of $\ln(m_H/k^\cut)$ they contain. Since the anomalous dimension contributions from clustering involve only the beam and soft functions, the total evolution factor from RG evolution for each veto observable is
\begin{align}
U_{\Delta}^{(2)} (\mu_S, \mu_J)
&= \exp \biggl[ \int_{\mu_S}^{\mu_J} \frac{\df\mu}{\mu} \biggl(\frac{\as(\mu)}{\pi}\biggr)^{\!2}  C_{\Tau}^{(2)} (\ln R) \biggr]
 , \nn \\
U_{\Delta}^{(2)} (\nu_S, \nu_J)
&= \exp \biggl[ \int_{\nu_S}^{\nu_J} \frac{\df\nu}{\nu} \biggl(\frac{\as(\mu)}{\pi}\biggr)^{\!2}  C_{p_T}^{(2)} (\ln R) \biggr]
.\end{align}
Although the logarithms of $R$ are exponentiated here, they are not connected with any scale in the effective theory. Below, we will show that the logarithms of $R$ indeed appear in the exponent, meaning the above procedure is correct.  However, the non-cusp anomalous dimension at $\ord{\as^n}$ contains a term $\sim \ln^{n-1}\! R$, so for small $R$ the perturbative series for the anomalous dimensions contains large unresummed logarithms. Equivalently, the perturbative series in the exponent is not resummed and contains terms $\alpha_s^n \ln^{n-1}\! R \ln(m_H/k^\cut)$. For $R\sim\lambda$ these are unresummed NLL corrections. Hence, the exponentiation of the $\ord{\as^2}$ clustering logarithms is doing nothing to tame the NLL clustering logarithms at higher orders. The complete resummation of the clustering logarithms would require to calculate the entire tower of coefficients simultaneously.

\subsection{Higher-Order Structure of Clustering Logarithms}

The higher-order clustering logarithms are simplest to examine in the soft function, and RG constraints can be used to relate them to the beam functions and determine the overall effect on the cross section. The soft function at $\ord{\as^n}$ can be written as
\begin{equation}
S^{(n)} (\pTcut) = \int\!\df \Phi_n \, \cA_n(\Phi_n) \, \cM_n(\Phi_n,\pTcut)
\,,\end{equation}
where $\Phi_n$ and $\cA_n$ are the $n$-particle phase space and matrix element.  Non-Abelian exponentiation~\cite{Gatheral:1983cz, Frenkel:1984pz, Gardi:2010rn, Mitov:2010rp} implies that the eikonal matrix elements exponentiate, and can be factorized into c-web contributions, where a c-web is a diagram connecting eikonal lines that cannot be separated into lower order c-webs by cutting each eikonal line once. This means that the amplitude can be written as
\begin{equation}
\cA_n (\Phi_n) = \sum_W N_W \biggl[ \prod_{w\in W} \cA_w (\Phi_{n_w}) \biggr]
\,,\end{equation}
where $W$ is a set of c-webs that partition the $n$-particle state, and $w$ is an individual c-web with $n_w$ particles.  $N_W$ is a combinatoric factor from exponentiation of the matrix elements and $\cA_w$ is the matrix element for the c-web $w$.

This factorization is useful because collinear singularities between final-state partons only exist when they are in the same c-web.  Therefore if two partons in different c-webs are clustered into the same jet then the rate is suppressed by the area of the jet in rapidity and azimuthal angle, which is $\ord{R^2}$.  At leading power, this implies that each jet's constituents are in the same c-web, and for a given set of c-webs $W$ the measurement function can be factorized over this set:
\begin{equation}
\cM_n (\Phi_n, \pTcut) = \prod_{w\in W} \cM_{n_w} (\Phi_{n_w})
\,.\end{equation}
Note that $\cM_{n_w}$ is the same measurement function as $\cM_n$ but over $\Phi_{n_w}$ instead of $\Phi_n$.  Using the exponentiation of matrix elements,
\begin{equation}
S(\pTcut) = \exp\biggl[ \sum_{\text{c-webs }w} S_w (\pTcut) \biggr]
\,,\end{equation}
where
\begin{equation}
S_w = \int \df \Phi_{n_w} \, \cA(\Phi_{n_w}) \, \cM_{n_w} (\Phi_{n_w}, \pTcut)
\,.\end{equation}

As argued in \app{RGconstraints}, RG invariance implies that the clustering logarithms can enter into at most the non-cusp anomalous dimension for the beam and soft functions at each order. For the $p_T$ jet veto, the clustering logarithms will be a part of the $\nu$ non-cusp anomalous dimension.

\begin{figure*}[ht!]
\centering
\includegraphics[width=\columnwidth]{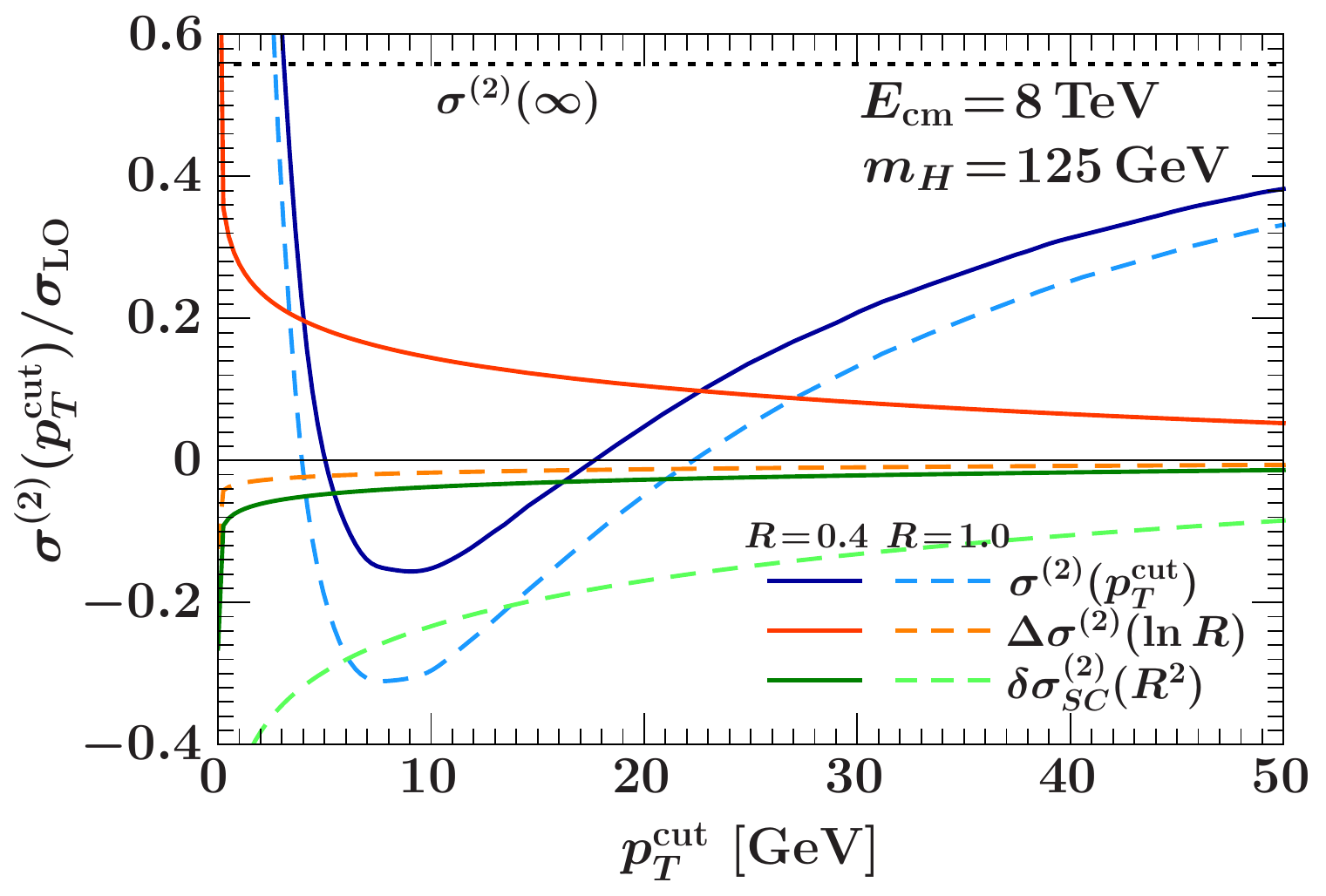}
\hfill%
\includegraphics[width=\columnwidth]{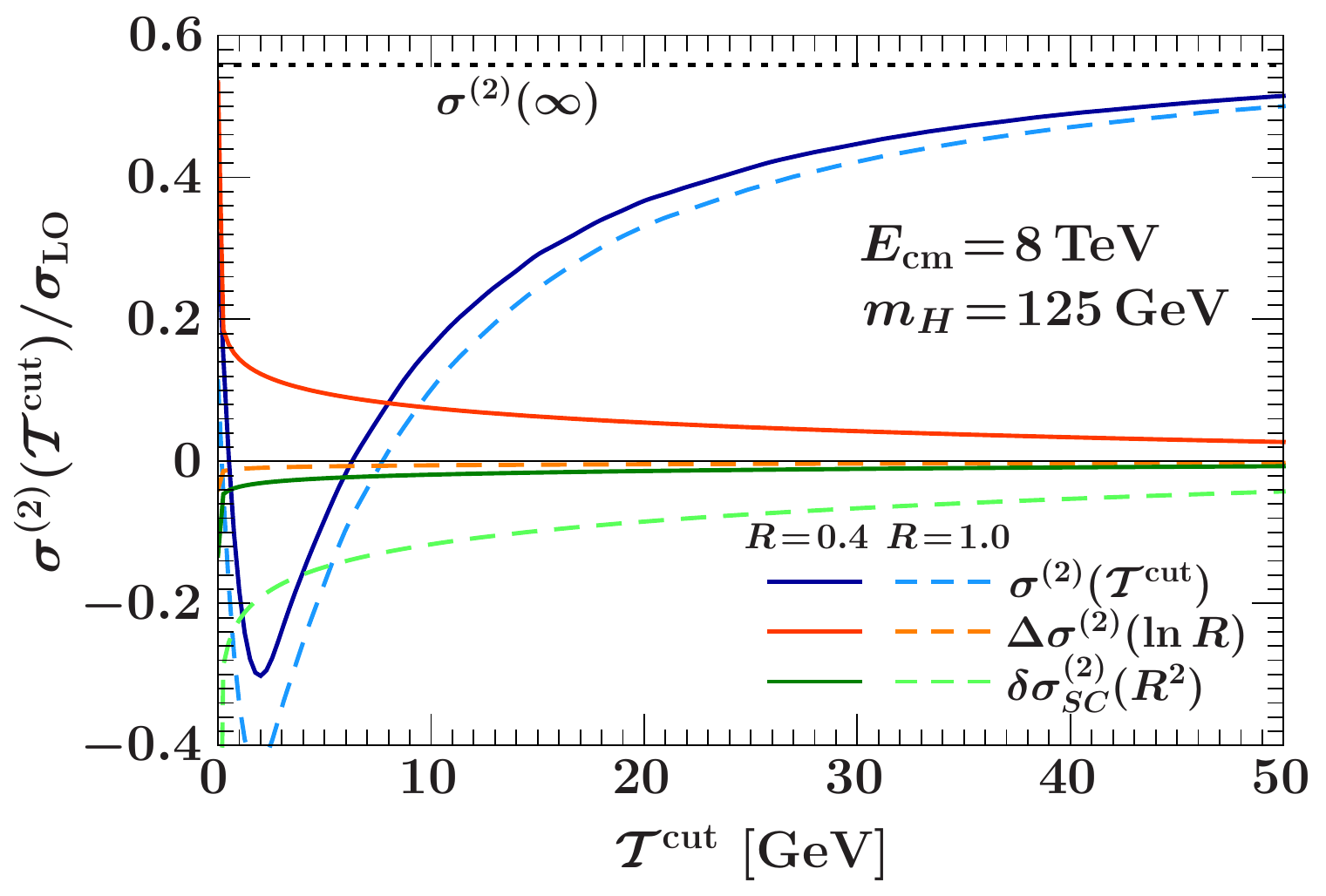}
\vspace{-0.5ex}
\caption{Numerical size of various $\ord{\as^2}$ terms for a cut on $\pTj$ (left panel) and on $\Tauj$ (right panel) for $m_H = 125\GeV$ at a $8 \TeV$ LHC.  The terms are scaled by the leading order cross section, $\sigma_{\rm LO}$. We show two values of the jet radius, $R = 0.4$ (solid curves) and $R = 1.0$ (dashed curves) for the $\kt$ class of jet algorithms.  Shown are the $\ord{\as^2}$ contribution to the total inclusive cross section with no veto, $\sigma^{(2)} (\infty)$, the full $\ord{\as^2}$ contribution with a jet veto, $\sigma^{(2)} (k^\cut)$ with $k^\cut = \pTcut$ or $\Taucut$, the $\ord{\as^2}$ clustering logarithm corrections, $\Delta \sigma^{(2)} (\ln R)$, and the soft-collinear mixing contribution, $\delta \sigma_{SC}^{(2)} (R^2)$. At small $R$, the clustering logarithms are sizable while the soft-collinear mixing terms are small. At large $R$ the situation is reversed.}
\label{fig:vetoplot}
\end{figure*}

Just as they did at $\ord{\as^2}$, collinear divergences between particles in the c-web will produce logarithms of $R$ as a finite artifact of the divergence.  Since there are at most $n_w-1$ collinear divergences in a c-web of $n_w$ particles, this implies there are at most $n_w - 1$ logarithms of R.  Therefore, the general form of the clustering logarithms in the soft function $S_w$ for the c-web is, in the $\Tau$ veto case,
\be
\Delta S_w (\Taucut) = \left(\frac{\as}{\pi}\right)^{n_w} C_{\Tau}^{(n_w)} (\ln R) \, \ln \frac{\mu}{\Taucut} \,,
\ee
and in the $p_T$ veto case,
\be
\Delta S_w (\pTcut) = \left(\frac{\as}{\pi}\right)^{n_w} C_{p_T}^{(n_w)} (\ln R) \, \ln \frac{\nu}{\pTcut} \,.
\ee
Both $C_{\Tau}^{(n_w)}$ and $C_{p_T}^{(n_w)}$ contain at most $n_w - 1$ powers of $\ln R$.  These give rise to non-cusp anomalous dimension contributions.  RG evolution of the soft function and beam function clustering logarithms from the c-web $w$ gives rise to an all-orders contribution to $\sigma$ of the form
\begin{align}
U_{\Delta}^{(n_w)} (\mu_S, \mu_J)
&= \exp \biggl[ \int_{\mu_S}^{\mu_J} \frac{\df\mu}{\mu} \biggl(\frac{\as(\mu)}{\pi}\biggr)^{n_w} C_{\Tau}^{(n_w)} (\ln R) \biggr]
\,, \nn \\
U_{\Delta}^{(n_w)} (\nu_S, \nu_J)
&= \exp \biggl[ \int_{\nu_S}^{\nu_J} \frac{\df\nu}{\nu} \biggl(\frac{\as(\mu)}{\pi}\biggr)^{n_w}  C_{p_T}^{(n_w)} (\ln R) \biggr] \,.
\end{align}
Since $C^{(n_w)}$ contains a $\ln^{n_w -1} R$ term, when $R\sim \lambda$ this series is NLL.  These terms are not directly obtained by resummation of lower order terms, underscoring the breakdown of resummation unless the coefficients of the clustering logarithms can all be calculated at once.

\section{Conclusions}
\label{sec:conclusions}

Jet vetoes are an important part of many experimental analyses. To obtain precise theoretical predictions, the resummation of large logarithms of the jet-veto scale is important, but they must account for effects of the jet algorithm. The algorithm generates two effects that can inhibit resummation, namely soft-collinear mixing and clustering logarithms. From a formal perspective, there are two scaling limits to consider, $R \sim \lambda$ and $R \gg \lambda$. Soft-collinear mixing arises at large values of the jet radius $R \gg\lambda$ from correlations between soft and collinear radiation into the same jet. This mixing affects the resummation of logarithms of the veto scale at NNLL and beyond and must be accounted for in order to prove an all-orders factorization theorem. When $R\sim \lambda$, soft-collinear mixing effects are power suppressed, but clustering logarithms are important, which are a remnant of collinear singularities between particles in the jet that cancel in the total cross section.  In the small $R$ limit, new clustering logarithms at NLL arise at each order and cannot currently be resummed. Therefore, there is a tradeoff between these effects depending on the value of $R$ chosen, and it is important to understand the relative sizes of these effects and incorporate them into estimates of theory uncertainties.

In \sec{factorization}, we contrasted the factorization picture for inclusive observables $\Tau_B$ and $E_T$ with the exclusive jet-based observables $\Tauj$ and $\pTj$.  The factorization and resummation for the inclusive observables are well understood. In this paper we have explored the effects of soft-collinear mixing and clustering on the factorization and resummation for the exclusive observables.  In \secs{scmixing}{clusteringlogs}, we calculated the soft-collinear mixing and clustering logarithms at $\ord{\as^2}$.

The resummation for a $\pTj$ veto was recently considered in Refs.~\cite{Banfi:2012yh, Becher:2012qa}.  In Ref.~\cite{Banfi:2012yh}, the resummation was performed to NLL, but the clustering logarithms were only included at fixed $\ord{\alpha_s^2}$. Our calculation of the leading clustering logarithms agree with the analytic results given in Ref.~\cite{Banfi:2012yh}. Reference~\cite{Becher:2012qa} works in the limit $\lambda \ll R\sim 1$. They argue that in this limit the factorization formula in \eq{VetopT_Fact} holds up to $\ord{\lambda}$ power corrections, and based on this extend the resummation to NNLL. While clustering logarithms do not pose a problem for $R\sim 1$, we disagree that the factorization formula in \eq{VetopT_Fact} holds in this limit. Since soft-collinear mixing contributions are in fact not power suppressed for $R\sim 1$, the factorization formula does not reproduce the full NNLL structure of the cross section at leading power. Therefore, the impact of these terms on the perturbative uncertainties found in Ref.~\cite{Becher:2012qa} should be examined.

We can investigate the numerical size of the jet-algorithm effects using Higgs production through gluon fusion as an example.  The cross section for a veto $k^\cut$ on $\Tauj$ or $\pTj$ can be expanded in fixed order as
\be
\sigma(k^\cut) = \sigma_{\rm LO} + \sigma^{(1)} (k^\cut) + \sigma^{(2)} (k^\cut) + \ord{\as^3} \,,
\ee
where $\sigma^{(n)}$ is $\ord{\as^n}$ relative to $\sigma_{\rm LO}$.  The $\ord{\as}$ terms, $\sigma^{(1)}$, do not depend on the jet algorithm, and hence are the same for $\Tauj$, $\pTj$ and $\Tau_B$, $E_T$, respectively.  The $\ord{\as^2}$ terms, $\sigma^{(2)}$, include the effect of clustering logarithms, $\Delta \sigma^{(2)} (\ln R)$, and soft-collinear mixing, $\delta \sigma^{(2)} (R^2)$,
\be
\sigma^{(2)} (k^\cut) \supset \Delta \sigma^{(2)} (\ln R) \,, \, \delta \sigma^{(2)} (R^2) \,.
\ee
In \fig{vetoplot}, we plot these terms in units of $\sigma_{\rm LO}$ as a function of $\pTcut$ (left panel) and $\Taucut$ (right panel), for $R = 0.4$ (solid curves) and $R = 1.0$ (dashed curves).  For comparison, we also show the full $\ord{\as^2}$ correction to the vetoed cross section, $\sigma^{(2)} (k^\cut)$, as well as the $\ord{\as^2}$ correction to the total cross section without a veto, $\sigma^{(2)} (k^\cut = \infty)$~\cite{Anastasiou:2004xq, Anastasiou:2005qj, Catani:2007vq, Grazzini:2008tf, Campbell:1999ah, Campbell:2010cz}.

The tradeoff between soft-collinear mixing and clustering logarithms is clear in \fig{vetoplot}.  For $R = 0.4$, the clustering logarithms dominate over the mixing terms.  For $R = 1.0$, the mixing terms dominate over the clustering logarithms.  In each case, the size of the numerically more important term is an appreciable fraction of the $\ord{\as^2}$ correction.  The difference between $\sigma^{(2)} (k^\cut)$ and $\sigma^{(2)} (\infty)$ primarily comes from the logarithmic terms, while the absolute size of $\sigma^{(2)} (k^\cut)$ relative to $\sigma^{(2)} (\infty)$ is indicative of the substantial cancellation between the jet-veto logarithms and the large NNLO $K$-factor for the total cross section~\cite{Stewart:2011cf}. Given that depending on $R$ the clustering or soft-collinear mixing terms are a sizable fraction of the logarithms $\ln(m_H/k^\cut)$ that one is trying to resum, it is crucial to understand their size at higher orders and correctly fold them into resummed uncertainty estimates.  This has not been done in previous studies of the Higgs cross section where resummation for the $p_T$ veto is performed.

The standard jet radii in Higgs analyses are $R = 0.4$ for ATLAS and $R = 0.5$ for CMS.  As seen in \fig{vetoplot}, with these values, at $\ord{\as^2}$ the clustering logarithms are numerically important and the mixing terms can be regarded as power corrections.  This suggests that the phenomenologically relevant limit is $R \sim \lambda$, where soft-collinear factorization can be applied but clustering logarithms should be regarded as NLL. To proceed one should study the impact of the clustering logarithms on the resummed perturbative series in order to properly take them into account in a resummed uncertainty estimate. For this purpose, a calculation of the $\ord{\as^3}$ clustering logarithms would provide very useful information.

\emph{Note added in proofs:} After this work was completed, Ref.~\cite{Banfi:2012jm} appeared, where a NNLL resummation formula for the cross section with a $p_T$ jet veto is presented for $R\sim 1$ that accounts for the contributions due to soft-collinear mixing. This formula is equivalent to modifying \eq{VetopT_Fact} to include these terms by hand, and is consistent with our analysis that \eq{VetopT_Fact} by itself is insufficient to perform the resummation beyond NLL when $R\sim 1$. (Ref.~\cite{Banfi:2012jm} does not discuss the case $R\sim \lambda$.) The structure of the $\ord{\as^2}$ soft-collinear mixing terms suggests that for the $p_T$ case one may be able to absorb them into the beam and soft functions in \eq{VetopT_Fact} by performing additional collinear zero-bin subtractions in the soft sector. It is an open question and beyond the scope of this work if such an approach can be extended to all orders in $\as$ such that \eq{VetopT_Fact} becomes a valid all-order factorization theorem.

\begin{acknowledgments}
We thank Andrea Banfi, Thomas Becher, and Matthias Neubert for discussions.  We thank Iain Stewart for comments on the manuscript.
J.W. and S.Z. thank the DESY Theory Group for hospitality while portions of this work were completed.  This work was supported in part by the DFG Emmy-Noether Grant No. TA 867/1-1, by the Director, Office of Science, Office of High Energy Physics of the U.S. Department of Energy under the Contract No. DE-AC02-05CH11231. J.W. was supported by the LHC Theory Initiative, under the National Science Foundation Grant No. PHY-0705682.
\end{acknowledgments}

\appendix

\section{Renormalization Group Constraints}
\label{app:RGconstraints}

The anomalous dimensions of the beam and soft functions for the four observables that we study are constrained by RG invariance.  We give a brief summary of the properties of the anomalous dimensions and the general form for each observable.  For the exclusive observables, we work in terms of the veto variable $\Taucut$ or $\pTcut$, and these forms are useful when discussing clustering logarithms in \sec{clusteringlogs}.

The hard function is universal for all the observables used, and the hard anomalous dimension is
\begin{equation}
\gamma_H^{gg} (m_H, \mu) = 2\Gcusp [\as(\mu)] \ln \frac{m_H^2}{\mu^2} + 2\gamma_H^g [\as(\mu)]
\,.\end{equation}
Consistency of the factorization theorem requires that the hard, beam, and soft anomalous dimensions cancel.  For the four observables, this consistency is slightly different.

For beam thrust, the consistency relation is
\begin{equation}
\gamma_H^{gg} (m_H, \mu) \delta(\Tau_B) + 2m_H\gamma_B^g (m_H\Tau_B,\mu) + \gamma_S^{gg} (\Tau_B,\mu) = 0
\,.\end{equation}
Analogously, for the $\Tauj < \Tau^\cut$ veto on jets, the consistency relation is
\begin{equation}
\gamma_H^{gg} (m_H, \mu) + 2\gamma_B^g (m_H\Taucut,\mu) + \gamma_S^{gg} (\Taucut,\mu) = 0
\,.\end{equation}
Both of these factorization theorems are described by \SCETI.  The scale dependence in the beam and soft functions is fixed by power counting; the beam function scale is $\mu_B^2 = m_H \Tau$ and the soft function scale is $\mu_S = \Tau$, where $\Tau = \Tau_B$ or $\Taucut$.  This constrains the beam and soft anomalous dimensions, so that the coefficients of the cusp anomalous dimension are fixed relative to that of the hard function.  For beam thrust, these anomalous dimensions are
\begin{align}
\gamma_B^g (t, \mu) &= - 2 \Gcusp [\as(\mu)]\, \frac{1}{\mu^2} \cL_0\Bigl(\frac{t}{\mu^2}\Bigr) + \gamma_B^g [\as(\mu)] \,\delta(t)
\,, \nn \\
\gamma_S^{gg} (k, \mu) &= 4 \Gcusp [\as(\mu)]\, \frac{1}{\mu} \cL_0\Bigl(\frac{k}{\mu}\Bigr) + \gamma_S^{gg} [\as(\mu)] \,\delta(k)
\,,\end{align}
where $\cL_0(x) = [\theta(x)/x]_+$ denotes the usual plus distribution. For a $\Tauj$ veto on jets, the anomalous dimensions are
\begin{align}
\gamma_B^g (m_H\Taucut, \mu)
&= - 2 \Gcusp [\as(\mu)] \ln\frac{m_H \Taucut}{\mu^2}
\nn \\ & \quad
+ \gamma_{B\,\rm jet}^g [\as(\mu)]
\,, \nn \\
\gamma_S^{gg} (\Taucut, \mu)
&= 4 \Gcusp [\as(\mu)] \ln\frac{\Taucut}{\mu}
\nn \\ & \quad
+ \gamma_{S\,\rm jet}^{gg} [\as(\mu)]
\,.\end{align}
The non-cusp anomalous dimensions for these two observables agree at $\ord{\as}$, but due to the different structure of their factorization theorems (beam thrust contains a convolution while the $\Tauj$ veto does not), the resummed distributions only coincide at leading logarithmic order.

For the $p_T$-based observables, which are described by \SCETII, rapidity divergences exist and must be separately regulated.  The rapidity renormalization group formalizes the regularization of rapidity divergences by introducing separate anomalous dimensions for the beam (or jet) and soft functions~\cite{Chiu:2011qc, Chiu:2012ir}.  This rapidity renormalization group works much like the traditional RGE, with a scale $\nu$ that is the analog to the usual $\mu$ scale.

Power counting and renormalization group invariance provide strong constraints on the anomalous dimensions for $p_T$-based observables.  Rapidity divergences are regulated by factors inserted into soft and collinear Wilson lines; in each sector the regulator and its scaling are
\begin{align}
\text{beam: } &  \; \nu^{\eta} \lvert \bn \cdot P_{g} \rvert^{-\eta} \sim \nu^{\eta} \bigl(\lambda^0 \bigr)^{-\eta}
\,, \nn \\
\text{soft: } & \; \nu^{\eta} \lvert P_{3g} \rvert^{-\eta} \sim \nu^{\eta} \bigl(\lambda \bigr)^{-\eta} \,,
\end{align}
where $P^{\mu}$ is the momentum operator, and the group momentum ($P_g$) of connected webs of gluons is regulated.  In each beam function the large component of momentum is regulated.  This power counting implies that in the rapidity RG the beam function scale is $\nu_B = m_H$ and the soft function scale $\nu_S = p_T$, where $p_T = E_T$ or $\pTcut$.  In addition, in  the standard RG in $\mu$, the beam and soft functions live at the low scale, $\mu_B = \mu_S = p_T$.

For the $p_T$-based observables, there are two consistency relations: one in $\mu$-space and one in $\nu$-space.  For the inclusive $E_T$ observable, the consistency relations are
\begin{align} \label{eq:ETconsistency}
0 &=  \gamma_H^{gg} (m_H, \mu) \delta(E_T) + 2\gamma_B^{\mu,g} (m_H, E_T,\mu,\nu)
\nn \\ & \quad
+ \gamma_S^{\mu,gg} (E_T,\mu,\nu)
\,, \nn \\
0 &= 2\gamma_B^{\nu,g} (E_T,\mu,\nu) + \gamma_S^{\nu,gg} (E_T,\mu,\nu)
\,.\end{align}
And for a $\pTj$ veto on jets, the consistency relations are
\begin{align}
0 &= \gamma_H^{gg} (m_H, \mu) + 2\gamma_B^{\mu,g} (m_H, \pTcut,\mu,\nu)
\nn \\ & \quad
+ \gamma_S^{\mu,gg} (\pTcut,\mu,\nu)
\,, \nn \\
0 &= 2\gamma_B^{\nu,g} (\pTcut,\mu,\nu) + \gamma_S^{\nu,gg} (\pTcut,\mu,\nu)
\,.\end{align}
The general forms of the beam and soft anomalous dimensions can be constrained through these relations as well as the exactness condition,
\begin{equation}
\frac{\df}{\df\ln \mu} \gamma_F^{\nu}(\mu, \nu) = \frac{\df}{\df\ln \nu} \gamma_F^{\mu}(\mu, \nu)
\,,\end{equation}
for $F = B,S$. We ignore running coupling effects in this exactness relation that will contribute to the $\nu$ anomalous dimensions starting at $\ord{\as^2}$. For $E_T$, the $\mu$ anomalous dimensions have the form
\begin{align}
&\gamma_B^{\mu,g} (m_H, E_T, \mu, \nu)
\nn\\ & \qquad
= \biggl\{ 2\Gcusp [\as(\mu)] \ln\frac{\nu}{m_H}
+ \gamma_B^{\mu,g} [\as(\mu)] \biggr\} \delta(E_T)
\nn\\ & \qquad\quad
+ \gamma^{\mu} [\as(\mu)]\, \frac{1}{\mu} \cL_0\Bigl(\frac{E_T}{\mu}\Bigr)
\,, \nn \\
&\gamma_S^{\mu,gg} (E_T, \mu, \nu)
\nn\\ & \qquad
= \biggl\{ 4\Gcusp [\as(\mu)] \ln\frac{\mu}{\nu} + \gamma_S^{\mu,gg} [\as(\mu)] \biggr\} \delta(E_T)
\nn \\ & \qquad\quad
 - 2\gamma^{\mu} [\as(\mu)]\, \frac{1}{\mu} \cL_0\Bigl(\frac{E_T}{\mu}\Bigr)
\,,\end{align}
and the $\nu$ anomalous dimensions have the form
\begin{align}
\gamma_B^{\nu,g} (E_T, \mu, \nu)
&= -2\Gcusp [\as(\mu)]\, \frac{1}{\mu} \cL_0\Bigl(\frac{E_T}{\mu}\Bigr)
\nn \\ & \quad
+ \gamma^{\nu,g} [\as(\mu)]\, \delta(E_T)
\,, \nn \\
\gamma_S^{\nu,gg} (E_T, \mu, \nu)
&= 4\Gcusp [\as(\mu)]\, \frac{1}{\mu} \cL_0\Bigl(\frac{E_T}{\mu}\Bigr)
\nn \\ & \quad
- 2\gamma^{\nu,g} [\as(\mu)]\, \delta(E_T)
\,.\end{align}
The cusp anomalous dimension dependence in the hard function fixes every part of the beam and soft anomalous dimensions except for the non-cusp terms and a cusp term $\gamma^{\mu} [\as(\mu)]$ in the beam and soft $\mu$ anomalous dimensions that cancels between them.  Because the beam and soft functions have the same $\mu$ scale, as long as they are RG evolved in $\mu$ in the same way this $\gamma^{\mu}$ term will cancel. At fixed-order, this cancellation is guaranteed and has no effect on the fixed-order logarithms.  It is absent in the anomalous dimensions at $\ord{\as}$, and may be absent at all orders.  Additionally, there is a constraint on the non-cusp anomalous dimensions from \eq{ETconsistency}, $2\gamma_H^g [\as] + 2\gamma_B^{\mu,g} [\as] + \gamma_S^{\mu,gg} [\as] = 0$.  If the beam and soft functions are RG evolved in $\mu$ in the same way, then it is irrelevant how the $\mu$ non-cusp anomalous dimensions are divided between them, since only the sum matters.

For the $\pTj$ veto the $\mu$ anomalous dimensions have the form
\begin{align}
\gamma_B^{\mu,g} (m_H, \pTcut, \mu, \nu)
&= 2\Gcusp [\as(\mu)] \ln\frac{\nu}{m_H}
\nn \\  & \quad
+ \gamma_{B\,\jet}^{\mu,g} [\as(\mu)]
+ \gamma^{\mu} [\as(\mu)] \ln\frac{\pTcut}{\mu}
\,, \nn \\
\gamma_S^{\mu,gg} (\pTcut, \mu, \nu)
& = 4\Gcusp [\as(\mu)] \ln \frac{\mu}{\nu}
+ \gamma_{S\,\jet}^{\mu,gg} [\as(\mu)]
\nn \\ & \quad
- 2\gamma^{\mu} [\as(\mu)] \ln\frac{\pTcut}{\mu}
\,,\end{align}
and the $\nu$ anomalous dimensions are
\begin{align}
\gamma_B^{\nu,g} ( \pTcut, \mu, \nu)
&= -2\Gcusp [\as(\mu)] \ln\frac{\pTcut}{\mu}
+ \gamma_{\jet}^{\nu,g} [\as(\mu)]
\,, \nn \\
\gamma_S^{\nu,gg} (\pTcut, \mu, \nu)
& = 4\Gcusp [\as(\mu)] \ln\frac{\pTcut}{\mu}
- 2\gamma_{\rm jet}^{\nu,g} [\as(\mu)]
\,.\end{align}
The same considerations as for the $E_T$ observable apply here. All the cusp parts are fixed by RG invariance with the hard function, and the only nontrivial unconstrained terms are the non-cusp $\nu$ anomalous dimensions of the beam and soft functions.

If clustering effects are associated with divergences in the beam and soft functions, then they can only impact the non-cusp beam and soft anomalous dimensions.  This implies that clustering corrections have at most a single logarithm of the veto variable at each order. However, as we saw in \sec{clusteringlogs}, they can produce additional logarithms of the jet radius $R$, because these are not associated with any scale and are therefore not constrained by the RGE.

\section{\boldmath Soft-Collinear Mixing at $\ord{\as^2}$}
\label{app:scmixingcalc}

In this appendix we present the $\ord{\as^2}$ calculations of the soft-collinear mixing contribution to the cross section, denoted as
$\delta \sigma^{\jet\,(2)}_{SC} = \delta \sigma^{\jet\,(2)}_{SC_a}+\delta \sigma^{\jet\,(2)}_{SC_b}$. As explained in \sec{scmixing}, this contribution arises from a correlation in the measurement between single independent emissions from the soft and collinear sectors, and formally breaks factorization if it is not power suppressed.  These mixing terms have the same form as the independent emission clustering terms at $\ord{\as^2}$ in the soft and collinear sectors, and the total $\ord{\alpha_s^2}$ clustering correction to the independent emission contribution to the cross section is given by [see \eq{Deltasigindep}]
\begin{equation}
\Delta \sigma^{\rm indep\,(2)} = \Delta \sigma^{\rm indep\,(2)}_{SS} + \Delta \sigma^{\rm indep\,(2)}_{CC}+\delta \sigma^{\jet\,(2)}_{SC}
\,.\end{equation}
The crucial result is that the soft-collinear mixing terms are NNLL terms that scale as $R^2$, requiring $R$ to scale as $\lambda$ for them to be formally power suppressed. We also calculate the $\ord{\alpha_s^2}$ clustering contribution from independent emissions in the soft function, $\Delta \sigma^{\rm indep\,(2)}_{SS}$.

The bare soft functions at $\ord{\as}$ and $\ord{\as^2}$ for independent emission in $\overline{\text{MS}}$ (without a measurement function) are
\begin{align}
S^{\rm b\, (1)}  &= \frac{\as C_A}{\pi} \frac{e^{\gamma_E \e}\mu^{2\e}}{\Gamma(1-\e)} \, \int_0^{\infty}\! \df k^+ \df k^-\, (k^+ k^-)^{-1-\e}
\,, \nn \\
S^{\rm b\, (2)}  &= \frac12 \bigl[S^{\rm b \, (1)} \bigr]^2
\,,\end{align}
where the label $\rm b$ stands for bare.  We have implemented the on-shell conditions and left only the $+,-$ light-cone components unintegrated.

The gluon beam function $B_g$ can be written as the convolution between a perturbative function $\cI (t, z,\mu)$ and the parton distribution function $f_g (x,\mu)$~\cite{Stewart:2010qs, Berger:2010xi},
\begin{equation} \label{eq:BgPDF}
B_g (t,x,\mu) = \!\sum_{j=\{g,q,\bar{q}\}} \int_x^1\! \frac{\df z}{z}\, \cI_{gj}(t, z,\mu)\, f_j \Bigl(\frac{x}{z},\mu\Bigr)
\,,\end{equation}
where $t =m_H k^+$ is the spacelike virtuality of the gluon entering in the hard interaction. The tree level, $B_g^{(0)}$, bare $\ord{\as}$ naive, $\widetilde B_g^{\rm b \, (1)}$, and zero-bin, $B_{g,0}^{\rm b \, (1)}$, beam functions (without a measurement function) are~\cite{Berger:2010xi}
\begin{align}
B_g^{(0)} &= \delta_{gj} \frac{1}{m_H}\delta(k^+) \delta(1-z)
\,,\nn \\
\widetilde B_g^{\rm b \, (1)}
&= \frac{\as C_A}{2\pi} \theta(z) \hat{P}_{gg} (z) \frac{1}{m_H} \frac{e^{\gamma_E \e}\mu^{2\e}}{\Gamma(1-\e)}
\int_0^{\infty}\! \df q^+ \df q^-
\nn \\ & \quad \times
(q^+)^{-1-\e} (q^-)^{-\e}\, \delta\Bigl(q^- - m_H \frac{1-z}{z}\Bigr)
\,,\nn \\
B_{g,0}^{\rm b \, (1)}
&= \frac{\as C_A}{\pi} \frac{1}{m_H}\delta(1-z) \frac{e^{\gamma_E \e} \mu^{2\e}}{\Gamma(1-\e)}
\nn \\ & \quad \times
\int_0^{\infty}\! \df q^+ \df q^-\, (q^+)^{-1-\e} (q^-)^{-1-\e}
\,,\nn \\
B_g^{\rm b \, (1)} &= \widetilde B_g^{\rm b \, (1)} - B_{g,0}^{\rm b \, (1)}
\,,\end{align}
where
\begin{equation}
\hat{P}_{gg}(z) = 2 \Bigl[ \frac{z}{1-z}+\frac{1-z}{z}+z(1-z) \Bigr]
\,.\end{equation}
We have dropped the quark contributions to the beam function as they will not produce logarithms in the mixing terms.  In $\widetilde B_g^{\rm b \, (1)}$, there is a singularity as $z\to1$ in the unregularized splitting function $\hat{P}_{gg}$ that is regulated by $q^-$.  When considering the full measurement (both the factorized and soft-collinear mixing terms), there is an IR divergence in $\cI$ that is canceled by the parton distribution functions; however, this cancellation occurs in the factorized measurement function and is not present in the mixing term.  The soft-collinear mixing correction has a single divergence and $B_g^{\rm b \, (1)}$ is proportional to $\delta(1-z)$, meaning there is no nontrivial convolution with the parton distribution functions.  Therefore we will just compute the coefficient of the leading order cross section, $\sigma_{\rm LO}$, which is given by
\begin{equation}\label{eq:sigmaLO}
\sigma_{\rm LO} = \sigma_0 \, H^{(0)}_{gg}(m_H,\mu) \int \df Y f_g(x_a,\mu)\, f_g(x_b,\mu)
\,,\end{equation}
with $x_{a,b} = (m_H / E_{\rm cm})e^{\pm Y}$.

\subsection{Clustering Effects for Independent Emissions in the Soft Function}

We perform the soft function calculations first, as the techniques and results will be used later in the soft-collinear mixing terms.  The independent emission terms are of the form
\begin{equation}
\Delta \sigma_{SS}^{\rm indep \, (2)} (k^\cut)
= S^{\rm b \, (2)} \int_0^{\pi}\! \frac{\df\Delta \phi}{\pi}\, \Delta \cM^{\rm jet \, (2)}_s
\,.\end{equation}
The integral over $\Delta \phi$ does not need to be regulated since there is no collinear divergence between the final-state particles.  The measurement function for a $\Tauj$ veto is
\begin{align}
\Delta \cM^{\rm jet \, (2)}_s (\Taucut)
&= 2\theta(\Delta R_{12} < R) \Bigl[ \theta(k_1^+ + k_2^+ < \Taucut)
\nn \\ &\quad
 - \theta(k_1^+ < \Taucut) \theta(k_2^+ < \Taucut) \Bigr]
 \,.\end{align}
We have multiplied by 2 to account for the case where the two gluons are both in the other hemisphere (where $\Tau$ is equal to the minus component of momenta), and dropped the $\ord{R^4}$ region of phase space where the two gluons are in different hemispheres but still cluster.  For a $\pTj$ veto, the measurement function is
\begin{align}
\Delta \cM^{\rm jet \, (2)}_s (\pTcut)
&= 2\theta(\Delta R_{12} < R) \Bigl[ \theta(p_{T1} + p_{T2} < \pTcut)
\nn \\ & \quad
- \theta(p_{T1} < \pTcut) \theta(p_{T2} < \pTcut) \Bigr]
 \,.\end{align}
In this case we drop the $\ord{R^4}$ correction from using the vector sum over transverse momenta in the combined constraint.

To evaluate the matrix element, we use the variables
\begin{align}
\Delta y &= \frac12 \ln \frac{k_1^- k_2^+}{k_1^+ k_2^-} \,, & y_t &= \frac14  \ln \frac{k_1^- k_2^-}{k_1^+ k_2^+} \,, \nn \\
\Taucut \, : \, \Tau_1 &= k_1^+ \,, & \Tau_2 &= k_2^+ \,, \nn \\
\pTcut \, : \, p_{T1} &= \sqrt{k_1^+ k_1^-} \,, & p_{T2} &= \sqrt{k_2^+ k_2^-} \,.
\end{align}
For the $\Tauj$ veto, the soft-function contribution is
\begin{align}
&\Delta \sigma_{SS}^{\rm indep \, (2)} (\Taucut)
\\ & \quad
= \frac{\sigma_{\rm LO}}{2} \Bigl(\frac{\as C_A}{\pi}\Bigr)^2  \frac{(e^{\gamma_E}\mu^2)^{2\e}}{\Gamma(1-\e)^2}\, 8\,
\int_{0}^{\infty}\! \df y_t \, e^{-4\e y_t}
\int_{-\infty}^{\infty}\! \df \Delta y
\nn \\ & \qquad \times
\int_0^{\pi}\! \frac{\df \Delta \phi}{\pi} \, \theta(\Delta R_{12} < R)
\int_0^{\infty} \! \df \Tau_1\, \df \Tau_2 \, (\Tau_1 \, \Tau_2)^{-1-2\e}
\nn \\ & \qquad \times
\Bigl[ \theta(\Tau_1 + \Tau_2 < \Taucut) - \theta( \Tau_1 < \Taucut) \theta( \Tau_2 < \Taucut) \Bigr]
\nn \,.\end{align}
Only the integral over $y_t$ produces a divergence, and the integrals are straightforward to evaluate.  The divergent term is
\begin{equation} \label{eq:DeltaSigSStau}
\Delta \sigma_{SS}^{\rm indep \, (2)}(\Taucut)
= -\sigma_{\rm LO} \frac{1}{\e} \Bigl(\frac{\as C_A}{\pi}\Bigr)^2 \Bigl(\frac{\mu}{\Taucut}\Bigr)^{4\e} \frac{\pi^2}{12} R^2
\,.\end{equation}
For $\pTj$, we need to regulate the rapidity divergences, for which we use the analytic regulator~\cite{Becher:2011dz}.
In this case, the regulator factor is
\begin{equation}
\nu^{2\alpha} (k_1^+ k_2^+)^{-\alpha} = \nu^{2\alpha} (p_{T1} p_{T2})^{-\alpha} e^{4\alpha y_t} \,.
\end{equation}
The amplitude and measurement function are independent of $y_t$. Hence, the soft function is proportional to
\begin{align}
\int_{-\infty}^{\infty}\! \df y_t \, e^{4\alpha y_t}
&= \int_{-\infty}^x\! \df y_t \, e^{4\alpha y_t} + \int_x^{\infty} \df y_t \, e^{4\alpha y_t}
\nn \\
&= \frac{1}{4\alpha} \bigl[ e^{4\alpha x} - e^{4\alpha x} \bigr] = 0
\,.\end{align}
We made the reason why this integral is zero explicit by breaking the range of integration into two parts: $(-\infty,x)$ and $(x,\infty)$.  In the lower range $\alpha > 0$ regulates the integral and in the upper range $\alpha < 0$ regulates the integral, and the two terms cancel. (This is precisely equivalent to the case of a scaleless integral in pure dimensional regularization with $\alpha$ here playing the role of $\e$.) Thus,
\begin{equation}
\Delta \sigma_{SS}^{\rm indep \, (2)} (\pTcut) = 0
\,.\end{equation}

\subsection{Soft-Collinear Mixing Terms}

The mixing term between soft and $n_a$-collinear emissions at $\ord{\as^2}$ is given by
\begin{equation} \label{eq:scmix}
(SC_a) = B_g^{\rm b \, (1)}S^{\rm b \, (1)}  \int_0^{\pi}\! \frac{\df \Delta\phi}{\pi}\, \delta \cM^{ {\rm jet} \, (2)}_{as}
\,,\end{equation}
and similarly for $(SC_b)$.  We have included the integral over the relative azimuthal angle $\Delta \phi$ between the soft and collinear particles. Since we are concerned with the leading divergences, we do not need to regulate $\phi$.  The $\ord{\as^2}$ measurement corrections for $\Tauj$ and $\pTj$, $\delta \cM^{ {\rm jet} \, (2)}(\Taucut)$ and $\delta \cM^{ {\rm jet}\,(2)} (\pTcut)$, are
\begin{align} \label{eq:scmixingmeas}
\delta \cM^{ {\rm jet} \, (2)}(\Taucut)
&= \theta(\Delta R_{sc} < R) \Bigl[ \theta(\Tau_c + \Tau_s < \Taucut)
\nn \\ & \quad
- \theta(\Tau_c < \Taucut)\, \theta(\Tau_s < \Taucut) \Bigr]
\,, \nn \\
\delta \cM^{ {\rm jet} \, (2)}(\pTcut)
&= \theta(\Delta R_{sc} < R) \Bigl[ \theta(p_{Tc} + p_{Ts} < \pTcut)
\nn \\ & \quad
- \theta(p_{Tc} < \pTcut)\, \theta(p_{Ts} < \pTcut) \Bigr]
\,.\end{align}

For the $\Tauj$ veto, the two soft-collinear mixing terms, $(SC_a)$ and $(SC_b)$, are equal.  They are made up of the naive and zero-bin beam function contributions, so that
\begin{equation}
(SC) = (S\widetilde{C}) - (SC^{(0)})
\,.\end{equation}

Changing variables from $k^-$ to $\Delta y$,
\begin{equation}
k^- = k^+ e^{2(y_c - \Delta y)} \,, \quad e^{-2y_c} = \frac{q^+ z}{m_H (1-z)}
\,.\end{equation}
the total correction from the naive beam function term in the mixing is
\begin{align}
&(S\widetilde{C}) (\Taucut)
\nn\\ & \quad
= 2\Bigl(\frac{\as C_A}{\pi}\Bigr)^2 \frac{(e^{\gamma_E }\mu^2)^{2\e}}{\Gamma(1-\e)^2}\,
\theta(z) \hat{P}_{gg} (z) \Bigl(\frac{z}{1-z}\Bigr)^{2\e} \frac{1}{m_H^{1+2\e}}
\nn \\ & \qquad \times
\int_{-\infty}^{\infty}\! \df \Delta y \, e^{2\e\Delta y} \int_0^{\pi}\! \frac{\df \Delta\phi}{\pi} \theta(\Delta R < R)
\nn\\ & \qquad \times
\int_0^{\infty}\! \df q^+ \df k^+\, (q^+ k^+)^{-1-2\e}
\\\nn & \qquad \times
\Bigl[ \theta(k^+\! + q^+\! < \Taucut) - \theta( q^+\! < \Taucut) \theta(k^+\! < \Taucut) \Bigr]
.\end{align}
The remaining integrals are finite, so we can set $\e = 0$ in the integrand after pulling out an overall scale dependence of $(\Taucut)^{-2\e}$.  The integrals over $\Delta y$ and $\Delta \phi$ give
\begin{equation}
\int_{-\infty}^{\infty}\! \df \Delta y \int_0^{\pi}\! \frac{\df\Delta \phi}{\pi} \, \theta(\Delta R < R) = \frac12 R^2
\,,\end{equation}
and the integrals over $q^+$ and $k^+$ give
\begin{align}
&\int_0^{\infty}\! \df q^+ \df k^+ \, \frac{1}{q^+ k^+} \Bigl[ \theta(k^+ + q^+ < \Taucut)
\\ & \qquad\qquad\qquad
- \theta( q^+ < \Taucut)\, \theta(k^+ < \Taucut) \Bigr]
= -\frac{\pi^2}{6}
\,. \nn\end{align}
Thus,
\begin{align}
(S\widetilde{C}) (\Taucut)
&= - \Bigl(\frac{\as C_A}{\pi}\Bigr)^2 \frac{\theta(z)}{m_H} \hat{P}_{gg} (z) \Bigl(\frac{z}{1-z}\Bigr)^{2\e}
\nn \\ & \quad \times
\biggl(\frac{\mu^2}{m_H \Taucut}\biggr)^{2\e} \frac{\pi^2}{6} R^2
.\end{align}
Expanding $z$ as a distribution about $z=1$ and keeping only the divergent term,
\begin{equation}
\theta(z) \hat{P}_{gg} (z) \Bigl(\frac{z}{1-z}\Bigr)^{2\e}
= -\frac{1}{\e} \delta(1-z) + \ord{\e^0}
\,,\end{equation}
the soft-collinear mixing term is
\begin{equation}
(S\widetilde{C}) (\Taucut)
= \frac{1}{\e} \Bigl(\frac{\as C_A}{\pi}\Bigr)^2 \, \frac{\delta(1-z)}{m_H}\, \biggl(\frac{\mu^2}{m_H \Taucut}\biggr)^{2\e} \frac{\pi^2}{6} R^2
\,.\end{equation}
The zero-bin contribution, $(SC^{(0)})$, comes from the soft limit of the naive contribution, which is obtained by taking the $z\to 1$ limit of $B_g^{\rm b \, (1)}$ in \eq{scmix}.  In this case, the correct scaling in the measurement function leaves it unchanged \cite{Jouttenus:2009ns}.  This leads to an unconstrained phase space for the clustered soft-collinear pair, meaning the zero-bin is proportional to a scaleless integral and vanishes,
\begin{equation}
(SC^{(0)}) (\Taucut) = 0 \,.
\end{equation}
Therefore, the total soft-collinear mixing contribution is\footnote{We thank Lorena Rothen for pointing out an error in our original calculation, which led us to reconsider the zero-bin contribution for $\Tauj$.}
\begin{align} \label{eq:DeltSigCStau}
\delta \sigma^{\rm jet \, (2)}_{SC}(\Taucut)
&= \sigma_{\rm LO} \frac{1}{\e} \Bigl(\frac{\as C_A}{\pi}\Bigr)^2 \biggl(\frac{\mu^2}{m_H \Taucut}\biggr)^{2\e} \frac{\pi^2}{6} R^2
\,. \end{align}

For the $\pTj$ veto, the zero-bin contribution vanishes for the same reason as the soft independent emission contribution. Hence, the soft-collinear mixing terms are just given by the naive contribution.  To evaluate the matrix element, we change variables from $q^+, k^+$ to $p_{Tc}, p_{Ts}$, where%
\begin{equation}
q^+ = \frac{p_{Tc}^2 z}{m_H (1-z)} \,, \quad k^+ = p_{Ts} \, e^{\Delta y} \, \frac{p_{Tc} z}{m_H (1-z)} \,.
\end{equation}
We also use the same change of variables from $k^-$ to $\Delta y$ as in the $\Tauj$ veto.  There are again rapidity divergences not regulated by $\e$ for which we use the analytic regulator.  The regulator factor for $(SC_a)$ is
\begin{equation}
\nu^{2\alpha} (q^+ k^+)^{-\alpha} = \nu^{2\alpha} (p_{Tc} p_{Ts})^{-\alpha} \left(\frac{p_{Tc} z}{m_H (1-z)}\right)^{-2\alpha}
\,.\end{equation}
The integrals over $\Delta y$ and $\Delta \phi$ can be performed as before, and we find
\begin{widetext}
\begin{align}
(SC_a) (\pTcut)
&= \Bigl(\frac{\as C_A}{\pi}\Bigr)^2 \frac{(e^{\gamma_E }\mu^2)^{2\e} }{\Gamma(1-\e)^2} \frac{1}{m_H}\, \theta(z) \hat{P}_{gg} (z) \Bigl(\frac{1-z}{z}\Bigr)^{2\alpha} (\nu m_H)^{2\alpha} \, R^2
\nn \\ & \quad \times
\int_0^{\infty}\! \df p_{Tc}\, \df p_{Ts}\, p_{Tc}^{-1-2\e-3\alpha} p_{Ts}^{-1-2\e-\alpha}
\Bigl[ \theta(p_{Tc} + p_{Ts} < \pTcut) - \theta( p_{Tc} < \pTcut)\, \theta( p_{Ts} < \pTcut) \Bigr]
 \,.\end{align}
\end{widetext}
We can set $\e = 0$ everywhere, as there is only a rapidity divergence as $z\to1$.  Performing the final integrals, we find
\begin{align}
(SC_a) (\pTcut)
&= -\Bigl(\frac{\as C_A}{\pi}\Bigr)^2 \frac{\theta(z) }{m_H}\,\hat{P}_{gg} (z) \Bigl(\frac{1-z}{z}\Bigr)^{2\alpha}
\nn\\* & \quad\times
\biggl( \frac{\nu m_H}{(\pTcut)^2} \biggr)^{2\alpha} \frac{\pi^2}{6} R^2
\,.\end{align}
Expanding in $z$ to extract the divergence, we get
\begin{align}
(SC_a) (\pTcut)
&= -\frac{1}{\alpha} \Bigl(\frac{\as C_A}{\pi}\Bigr)^2 \frac{\delta(1-z)}{m_H} \biggl( \frac{\nu m_H}{(\pTcut)^2} \biggr)^{2\alpha}
\frac{\pi^2}{6} R^2
\,.\end{align}
This gives the following correction to the cross section
\begin{equation}
\delta \sigma^{\rm jet \,(2)}_{SC_a}(\pTcut)
= - \sigma_{\rm LO} \frac{1}{\alpha} \Bigl(\frac{\as C_A}{\pi}\Bigr)^2 \left( \frac{\nu m_H}{(\pTcut)^2} \right)^{2\alpha} \frac{\pi^2}{6} R^2\,.
\end{equation}
The regulator for the mixing between soft and $n_b$-collinear sectors gives a difference scale dependence.  In this case the label component of collinear momentum is regulated, which effectively amounts to regulating the minus momentum component in the above calculation.  This changes the scale dependence to $(\nu/m_H)^{2\alpha}$ with an overall minus sign:
\begin{equation}
\biggl( \frac{\nu m_H}{(\pTcut)^2} \biggr)^{2\alpha} \to -(\nu/m_H)^{2\alpha}
\,.\end{equation}
Thus the entire soft-collinear mixing contribution to the cross section, $\delta \sigma^\jet_{SC_a}+\delta \sigma^\jet_{SC_b}$, is
\begin{align} \label{eq:DeltSigCSpT}
\delta \sigma^{\rm jet \, (2)}_{SC} (\pTcut)
&= -\sigma_{\rm LO} \frac{1}{\alpha} \Bigl(\frac{\as C_A}{\pi}\Bigr)^2 \frac{\pi^2}{6} R^2
\nn\\ & \quad\times
\biggl[ \biggl( \frac{\nu m_H}{(\pTcut)^2} \biggr)^{2\alpha} - \biggl(\frac{\nu}{m_H}\biggr)^{2\alpha} \biggr]
\nn\\
&= -\sigma_{\rm LO} \Bigl(\frac{\as C_A}{\pi}\Bigr)^2 \frac{2\pi^2}{3} R^2 \,\ln \frac{m_H}{\pTcut}
\,.\end{align}

\section{\boldmath Clustering Logarithms in the Soft Function at $\ord{\as^2}$}
\label{app:clusteringlogcalc}

Clustering logarithms first occur at $\ord{\as^2}$, and they are easiest to calculate in the soft function.  RG invariance can be used to extract the beam function contributions, as explained in \sec{clusteringlogs}.

The definition of what is a clustering effect is subtle because one must define what the effect is \emph{relative} to.  One ostensibly natural option is to define it relative to the cross section if no clustering takes place.  However, as explained in \sec{clusteringlogs}, this measurement is infrared unsafe due to collinear singularities between partons.  A more sensible ``primary'' measurement are the inclusive $\Tau_B$ or $E_T$ measurements.  These measurements are IR safe, and when a set of particles becomes collinear their contribution to the primary measurement is the same as the contribution to the observable that is vetoed (the $p_T$ or $\Tau$ of the jet). Here, we only keep the divergent term with the appropriate $\pTcut$ and $\Taucut$ scale dependence, as this is the piece connected to the soft and beam functions.  The finite terms do not take part in the resummation and can be captured as usual by matching the resummed result to the full fixed-order result at NNLO.

The $\ord{\as^2}$ soft measurement function for the clustering correction relative to the inclusive measurements is [see \eq{DeltaM2}]
\begin{align}
\nn\\
&\Delta \cM_s^{\jet\, (2)} (k^\cut)
\nn\\ & \qquad
= \biggl\{\theta(\Delta R_{12} < R) \theta(k_\jet < k^\cut)
\nn \\ & \qquad\quad
+ \theta(\Delta R_{12} > R) \theta(k_1 < k^\cut) \theta(k_2 < k^\cut) \biggr\}
\nn \\
& \qquad\quad
- \theta (k_1 + k_2 < k^\cut)
\,,\end{align}
where $k = \Tau$ or $p_T$.  In this measurement function, $k_\jet$ is the observable for the clustered pair.  For $k = \Tau$, $\Tau_\jet = \Tau_1 + \Tau_2$ except when the jet spans the boundary at the Higgs rapidity.  For $k = p_T$, the scalar sum of transverse momenta is not the same as the magnitude of the vector sum.  However, when $R \sim \lambda$, these differences become power-suppressed for both the $\Tau$ and $p_T$ observables.  Since this is the limit we are working in, and we are neglecting power-suppressed terms in this limit, the measurement functions for the clustering effect simplify to
\begin{align}
\Delta \cM^{\jet\,(2)}(k^\cut)
&= \theta(\Delta R_{12}\! > R) \Bigl[ \theta(k_1\! < k^\cut) \theta(k_2\! < k^\cut)
\nn\\ & \quad
- \theta (k_1 + k_2 < k^\cut) \Bigr]
\,.\end{align}
The phase space constraint on $\Delta R_{12}$ suggests a convenient set of coordinates for the calculation.  In terms of the rapidity $y_i$, azimuthal angle $\phi_i$, and observable $k_i$ (where $k = \Tau$ or $p_T$), the coordinates we use are
\begin{align} \label{eq:cluscalcvariables}
y_t &= \frac12 (y_1 + y_2) \,, \qquad \phi_t = \frac12 (\phi_1 + \phi_2)
\,, \nn \\
\Delta y &= y_1 - y_2 \,, \qquad \quad \Delta \phi = \phi_1 - \phi_2
\,, \nn \\
k_t &= k_1 + k_2 \,, \qquad \qquad\! z = \frac{k_1}{k_t}
\,.\end{align}
The full $\ord{\as^2}$ soft function matrix elements can be found in Ref.~\cite{Hornig:2011iu}. Using $k = p_T$, in terms of these coordinates the non-Abelian soft matrix elements are
\begin{widetext}
\begin{align}
\cA_A &= 4g^4 C_A^2  \, \frac{1}{p_{Tt}^4 \, z^2 (1-z)^2} \, \frac{1}{\cosh \Delta y - \cos\Delta\phi} \, \frac{1}{z^2 + (1-z)^2 + 2z(1-z) \cosh \Delta y} \nn \\
& \qquad \qquad \times \biggl\{ (z^2 + (1-z)^2) \cos \Delta\phi - z(1-z)(1 - \cos\Delta\phi \cosh\Delta y)
\nn \\
& \qquad \qquad \qquad + (1-\e) \frac{z^2 (1-z)^2 \sinh^2 \Delta y }{(\cosh \Delta y - \cos\Delta\phi) (z^2 + (1-z)^2 + 2 z(1-z) \cosh \Delta y)} \biggr\}
\,,\nn \\
\cA_f &= 4g^4 C_A T_R n_f \, \frac{1}{p_{Tt}^4 \, z^2 (1-z)^2} \, \frac{1}{(\cosh \Delta y - \cos\Delta\phi)} \, \frac{1}{(z^2 + (1-z)^2 + 2z(1-z) \cosh \Delta y)}
\nn \\
& \qquad \qquad \qquad \times \biggl\{ z(1-z) - \frac{2z^2 (1-z)^2 \sinh^2 \Delta y }{(\cosh \Delta y - \cos\Delta\phi) (z^2 + (1-z)^2 + 2 z(1-z) \cosh \Delta y)} \biggr\}
\,.\end{align}
In the small $\Delta R$ limit, the matrix elements simplify to
\begin{align} \label{eq:AsmallR}
\cA_A^R &= 4g^4 C_A^2 \, \frac{1}{p_{Tt}^4 z^2 (1-z)^2} \frac{2}{\Delta R^2} \biggl[ z^2 + (1-z)^2 + \frac{2z^2 (1-z)^2 \Delta y^2}{\Delta R^2} \biggr]
\,, \nn \\
\cA_f^R &= 4g^4 C_A T_R n_f \, \frac{1}{p_{Tt}^4 z^2 (1-z)^2} \frac{2}{\Delta R^2} \biggl[ z(1-z) - \frac{4z^2 (1-z)^2 \Delta y^2}{\Delta R^2} \biggr]
\,.\end{align}

We start with the calculation for $\pTj$. The measurement function can be written as
\begin{equation}
\Delta \cM^\jet_s(\pTcut) = \theta(\Delta R > R) \, \theta\biggl[\pTcut < p_{Tt} < \pTcut \frac{1}{\max(z,1-z)} \biggr]
\,.\end{equation}
The matrix element and measurement function are independent of $y_t$, meaning the integral over $y_t$ is unregulated.  To regulate this rapidity divergence we use the rapidity regulator, which regulates the $z$-component of the group momentum for the c-web through the factor
\begin{equation}
\nu^{\eta} \lvert 2 P_{3g} \rvert^{-\eta}
= \nu^{\eta} \, p_{Tt}^{-\eta} \, \Bigl\lvert 2z \sinh \Bigl(y_t + \frac12 \Delta y \Bigr)
 + 2(1-z) \sinh \Bigl(y_t - \frac12 \Delta y\Bigr) \Bigr\rvert^{-\eta}
\,.\end{equation}
Integrating over $y_t$ then gives a single $1/\eta$ divergence and this is the only divergent part of the calculation,
\begin{equation}
\nu^{\eta} p_{Tt}^{-\eta} \int_{-\infty}^{\infty}\! \df y_t \,
\Bigl\lvert 2z \sinh \Bigl(y_t + \frac12 \Delta y \Bigr)
 + 2(1-z) \sinh \Bigl(y_t - \frac12 \Delta y\Bigr) \Bigr\rvert^{-\eta}
= \frac{2}{\eta} \nu^{\eta} p_{Tt}^{-\eta} + \ord{\eta^0}
\,.\end{equation}
The matrix element scales simply with $p_{Tt}$ since it is the only dimensionful variable, and so $p_{Tt}$ can be easily integrated against the measurement function including the regulator factor.  This integral is
\begin{equation}
\int_0^{\infty}\! \df p_{Tt} \, p_{Tt}^{-1-4\e-\eta} \, \theta\biggl[\pTcut < p_{Tt} < \pTcut \frac{1}{\max(z,1-z)} \biggr]
= -(\pTcut)^{- 4\e - \eta} \ln[\max(z,1-z)]
\,.\end{equation}
Carrying through the integrals over the on-shell conditions, the soft function correction for the $p_T$ veto is
\begin{align}
\Delta S^{\rm b \, (2)} (\pTcut)
&= - \frac{8}{\eta}\,\frac{1}{(4\pi)^4} \biggl(\frac{\nu}{\pTcut}\biggr)^{\eta}
\!\int_0^1\! \df z \int_{-\infty}^{\infty}\!\!\! \df\Delta y \int_0^{\pi}\! \frac{\df\Delta \phi}{\pi} \, \frac{\ln[\max(z,1-z)]}{z(1-z)}\,
\theta(\Delta R > R) \Bigl[ p_{Tt}^4 z^2 (1-z)^2 \cA (z,\Delta y, \Delta \phi) \Bigr]
\,.\end{align}
The remaining integrals are finite and have the form
\begin{equation}
a \ln R + b + \ord{R}
\,.\end{equation}
We determine $a$ analytically and extract $b$ numerically.  To determine the coefficient of $\ln R$, we rewrite the matrix element as the difference
\begin{equation}
\cA = (\cA - \cA^R) + \cA^R
\,,\end{equation}
where $\cA^R$ is the matrix element expanded in the small $R$ limit, given in \eq{AsmallR}.  The difference $\cA - \cA^R$ is finite as $R\to 0$, meaning that up to $\ord{R^2}$ corrections,
\begin{equation}
\int\! \df\Delta y \df\Delta \phi \, \theta(\Delta R > R) \, (\cA - \cA^R)
= \int \df\Delta y \,\df\Delta \phi \, (\cA - \cA^R) + \ord{R^2}
\,.\end{equation}
Integrating the matrix element $\cA^R$ is simple, using the relations
\begin{align}
\int_{-\infty}^{\infty}\! \df\Delta y \int_0^{\pi} \frac{\df \Delta \phi}{\pi} \, \frac{1}{\Delta R^2} \, \theta(\Delta R > R) &= - \ln R + \ln 2\pi + \ord{R}
\,, \nn \\
\int_{-\infty}^{\infty} d\Delta y \int_0^{\pi} \frac{d\Delta \phi}{\pi} \, \frac{2\Delta y^2}{\Delta R^4} \, \theta(\Delta R > R) &= - \ln R + \ln 2\pi + \frac12 + \ord{R}
\,.\end{align}
Carrying through the integrals, we obtain
\begin{align}
&\int_0^1\! \df z \int_{-\infty}^{\infty}\! \df \Delta y \int_0^{\pi}\! \frac{\df \Delta \phi}{\pi} \frac{\ln[\max(z,1-z)]}{z(1-z)} \, \Big[ p_{Tt}^4 z^2 (1-z)^2 \cA_A^R \Big] \theta(\Delta R > R)
\nn \\ &\qquad
= 4g^4 C_A^2 \Bigl\{ -\frac{1}{36} (131 - 12\pi^2 - 132\ln 2) \ln R + \frac{1}{72} \bigl[ -13 + 12\ln 2 + 2\ln (2\pi)(131 - 12\pi^2 - 132\ln 2) \bigr] \Bigr\} \,, \nn \\
&\int_0^1\! \df z \int_{-\infty}^{\infty}\! \df \Delta y \int_0^{\pi}\! \frac{\df\Delta \phi}{\pi} \frac{\ln[\max(z,1-z)]}{z(1-z)} \Big[ p_{Tt}^4 z^2 (1-z)^2 \cA_f^R \Big] \theta(\Delta R > R)
\nn \\ & \qquad
= 4g^4 C_A T_R n_f \Bigl\{ \frac{1}{18} (23 - 24\ln 2) \ln R + \frac{1}{36} \bigl[ 13 - 12\ln 2 - 2\ln (2\pi)(23 - 24\ln 2) \bigr] \Bigr\}
\,.\end{align}
Performing the full integrals numerically, we get
\begin{align}
\int_0^1\! \df z\, \int_{-\infty}^{\infty} \df\Delta y \int_0^{\pi}\! \frac{\df\Delta \phi}{\pi} \frac{\ln[\max(z,1-z)]}{z(1-z)} \Bigl[ p_{Tt}^4 z^2 (1-z)^2 (\cA_A - \cA_A^R) \Bigr]
&= 4g^4 C_A^2 \, (4.66)
\,, \nn \\
\int_0^1\! \df z\, \int_{-\infty}^{\infty} \df\Delta y \int_0^{\pi}\! \frac{\df\Delta \phi}{\pi} \frac{\ln[\max(z,1-z)]}{z(1-z)} \Bigl[ p_{Tt}^4 z^2 (1-z)^2 (\cA_f - \cA_f^R) \Bigr]
&= 4g^4 C_A T_R n_f \, (0.138)
\,.\end{align}
Thus, the final result for the leading divergent corrections due to clustering in the bare soft function is
\begin{align} \label{eq:softfunctioncluspT}
\Delta S^{\rm b \, (2)} (\pTcut) 
&= \frac{1}{\eta} \, \Bigl(\frac{\as}{\pi}\Bigr)^2 \biggl(\frac{\nu}{\pTcut} \biggr)^{\eta}
\Bigl\{ C_A^2 \Bigl[ \frac{1}{18} (131 - 12\pi^2 - 132\ln 2) \ln R - 1.12 \Bigr]
\nn \\ & \quad
+ C_A T_R n_f \Bigl[ -\frac{1}{9} ( 23 - 24 \ln 2 ) \ln R + 0.764 \Bigr] \Bigr\}
\,.\end{align}
Note that the constant terms, $-1.12$ and $+0.764$, depend on the choice of inclusive observable that the clustering effect is defined relative to. Hence, they are different from those found in Ref.~\cite{Banfi:2012yh}, since here we use $E_T = \abs{\vec{p}_{T1}} + \abs{\vec{p}_{T2}}$ while Ref.~\cite{Banfi:2012yh} uses the vector sum $\abs{\vec{p}_{T1} + \vec{p}_{T2}}$. When using the latter in our calculation we reproduce the results in Ref.~\cite{Banfi:2012yh}.

The steps to calculate the clustering logarithms for $\Tauj$ proceed analogously.  The calculation is slightly more tedious due to the fact that a particle's contribution to the observable changes depending on what hemisphere it is in.  Therefore, there are two regions of rapidities $y_{1,2}$ of the two particles to consider:
\begin{enumerate}
\item[i)] $y_1, y_2 > 0$ and $y_1, y_2 < 0$: Both particles are in the same hemisphere.  These configurations will contribute to the divergent terms.
\item[ii)] $y_1 > 0, y_2 < 0$ and $y_1 < 0, y_2 > 0$: The particles are in opposite hemispheres. This region does not contribute to the divergent terms. The region of phase space where particles can cluster scales as $\ord{R^4}$ and can be neglected.
\end{enumerate}
The variables in \eq{cluscalcvariables} are also useful for this calculation.  Carrying through the calculation, we find the divergent terms contributing to the bare soft function
\begin{align} \label{eq:softfunctionclusTau}
\Delta S^{\rm b \, (2)} (\Taucut)
&= \frac{1}{4\e} \Bigl(\frac{\as}{\pi}\Bigr)^2 \biggl(\frac{\mu}{\Taucut} \biggr)^{4\e}
\Bigl\{ C_A^2 \Bigl[ \frac{1}{18} \big(131 - 12\pi^2 - 132\ln 2\big) \ln R - 0.937 \Bigr]
\nn \\ & \quad
+ C_A T_R n_f \Bigl[ -\frac{1}{9}\big( 23 - 24 \ln 2 \big) \ln R + 0.747 \Bigr] \Bigr\}
\,.\end{align}
Note that the divergent $\ln R$ terms are the same as for $\pTj$, while the constant terms differ.

\end{widetext}

\bibliographystyle{../physrev4}
\bibliography{../jets}

\end{document}